%% file: main.tex
\documentclass[conference]{IEEEtran}
\IEEEoverridecommandlockouts
\usepackage{cite}
\usepackage{amsmath,amssymb,amsfonts}
\usepackage{algorithmic}
\usepackage{graphicx}
\usepackage{textcomp}
\def\BibTeX{{\rm B\kern-.05em{\sc i\kern-.025em b}\kern-.08em
    T\kern-.1667em\lower.7ex\hbox{E}\kern-.125emX}}

\usepackage{tabularx}
\usepackage{placeins} 
\usepackage{tikz}
\usepackage{amsmath}
\usepackage{pifont}
\usepackage{booktabs} 
\usepackage{graphicx} 
\usepackage{wasysym}
\usepackage[table]{xcolor}
\usepackage[ruled,vlined]{algorithm2e}
\usepackage{hyperref}
\usetikzlibrary{positioning,backgrounds,fit,calc,arrows.meta}
\usepackage{multirow} 
\usepackage[caption=false,font=footnotesize]{subfig}
\usepackage[most]{tcolorbox}
\usepackage{paralist}
\usepackage[normalem]{ulem}

\usepackage{arydshln}     % for \hdashline
\usepackage{graphicx} % 用于 \resizebox

\newcommand{\hlhref}[2]{\href{#1}{\textcolor{black}{{#2}}}}
\newcommand{\heading}[1]{{\vspace{2pt}\noindent{\textbf{#1}}}}

\newtcolorbox{semibox}{
  enhanced,
  width=0.46\textwidth, 
  colback=gray!6,           % 淡灰背景
  colframe=black!50,        % 框线颜色
  boxrule=0.4pt,            % 框线粗细
  arc=2pt,                  % 圆角
  left=2pt,right=2pt,top=1pt,bottom=1pt,
  borderline west={1.0pt}{0pt}{black!70}, % 左侧加深
  borderline east={1.0pt}{0pt}{black!70}, % 右侧加深
  drop shadow                % 轻微投影，显得立体
}
\usepackage[most]{tcolorbox} % 导言区
\tcbset{colframe=black!50, colback=gray!10, boxrule=0.5pt, arc=2pt, boxsep=5pt, left=3pt, right=3pt, top=3pt, bottom=3pt, sharp corners}

\usepackage{booktabs}

\usepackage{fontawesome} % 在导言区引入

\usepackage{extarrows}
\usepackage{bm} % 导言区

\usepackage{tikz}
\usetikzlibrary{positioning}

\usepackage{makecell,array}
\usepackage{booktabs,multirow}
\usepackage{tikz}
\newcommand{\redx}{{\textcolor{red}{\ding{55}}}}
\newcommand{\greencheck}{{\textcolor{blue}{\ding{51}}}}
\usepackage{tikz}
\usepackage{tikz}
\newcommand*\circled[1]{\tikz[baseline=(char.base)]{
    \node[shape=circle,draw,line width=0.65pt,inner sep=0pt,minimum size=1em] (char) {\footnotesize #1};}}
\DeclareRobustCommand*\circledc[1]{\tikz[baseline=(char.base)]{
    \node[shape=circle,draw,line width=0.6pt,inner sep=0pt,minimum size=0.9em] (char) {\footnotesize #1};}}

\usepackage{filecontents}
\newcommand{\techName}[1]{\textsc{FlowShield}}

\newenvironment{packeditemize}{
	\begin{list}{$\bullet$}{
			\setlength{\labelwidth}{4pt}
			\setlength{\itemsep}{0pt}
			\setlength{\leftmargin}{\labelwidth}
			\addtolength{\leftmargin}{\labelsep}
			\setlength{\parindent}{0pt}
			\setlength{\listparindent}{\parindent}
			\setlength{\parsep}{0pt}
			\setlength{\topsep}{1pt}}}{\end{list}}

\begin{document}

\title{\textsc{FlowShield}: cryptocurrency anti-money laundering with transaction semantics parsing and fund flow tracking}

\makeatletter
\newcommand{\linebreakand}{%
  \end{@IEEEauthorhalign}
  \hfill\mbox{}\par
  \mbox{}\hfill\begin{@IEEEauthorhalign}
}

\author{
\IEEEauthorblockN{ Qishuang Fu}
\IEEEauthorblockA{
  %\textit{Faculty of Information Technology} \\
\textit{Monash University}\\
% Melbourne, Australia \\
qishuang.fu@monash.edu}
\and
\IEEEauthorblockN{ Andreas Deppeler}
\IEEEauthorblockA{
  %\textit{School of Business} \\
\textit{Monash University}\\
% Sunway, Malaysia \\
andreas.deppeler@monash.edu}
\and
\IEEEauthorblockN{ Joseph K. Liu}
\IEEEauthorblockA{
  %\textit{Faculty of Information Technology} \\
\textit{Monash University}\\
% Melbourne, Australia \\
joseph.liu@monash.edu}
\and
\IEEEauthorblockN{ Yixin Liu}
\IEEEauthorblockA{
  %\textit{School of ICT} \\
\textit{Griffith University}\\
% Brisbane, Australia \\
yixin.liu@griffith.edu.au}
\linebreakand
\IEEEauthorblockN{Shirui Pan}
\IEEEauthorblockA{
  %\textit{School of ICT} \\
\textit{Griffith University}\\
% Gold Coast, Australia \\
s.pan@griffith.edu.au}
\and
\IEEEauthorblockN{Qin Wang}
\IEEEauthorblockA{
  %\textit{CSIRO Data61} \\
\textit{CSIRO Data61}\\
% Sydney, Australia \\
qinwangtech@gmail.com}
\and
\IEEEauthorblockN{ Weiqing Wang}
\IEEEauthorblockA{
% \textit{Faculty of Information Technology} \\
\textit{Monash University}\\
% Melbourne, Australia \\
teresa.wang@monash.edu}
\and
\IEEEauthorblockN{ Tsz Hon Yuen}
\IEEEauthorblockA{
  %\textit{Faculty of Information Technology} \\
\textit{Monash University}\\
% Melbourne, Australia \\
john.tszhonyuen@monash.edu}
\thanks{Accepted to the 2026 IEEE International Conference on Data Mining (ICDM 2026).}
}
% %
% %%
% %% By default, the full list of authors will be used in the page
% %% headers. Often, this list is too long, and will overlap
% %% other information printed in the page headers. This command allows
% %% the author to define a more concise list
% %% of authors' names for this purpose.
% \renewcommand{\shortauthors}{Fu et al.}

% IEEE author block intentionally left commented out.
% \author{\IEEEauthorblockN{Qishuang Fu, Andreas Deppeler, Joseph K. Liu, Yixin Liu, Shirui Pan, Qin Wang, Weiqing Wang, and Tsz Hon Yuen}
% \IEEEauthorblockA{Affiliations omitted.}
% }

\maketitle

\begin{abstract}

%AML --> existing work --> our work short 

Cryptocurrency anti-money laundering (Crypto AML) is increasingly challenged by sophisticated laundering behaviors that rapidly fragment stolen assets through diverse semantics and across multiple blockchains.
Existing Crypto AML methods often simplify transaction semantics, rely on topology-centric signals, or output isolated detection labels.
In this paper, we present \textsc{FlowShield}, a Crypto AML framework for transaction-level laundering detection and investigator-facing report generation. 
\textsc{FlowShield} first recovers behavior-level semantics from observable relations, making laundering intents explicit. 
To trace value provenance and redistribution, \textsc{FlowShield} reconstructs fund-flow subgraphs from three complementary perspectives.
It then employs a text--structure fusion mechanism, enabling the interplay between large language model (LLM)-encoded semantics and flow texts with graph convolutional network (GCN)-encoded structure.
Beyond mere detection, \textsc{FlowShield} further generates readable suspicious activity reports (SARs), offering investigators concise summaries and explainable red flags. 
To address the data scarcity in multi-chain detection, we construct and open-source \textit{BybitML}, the first public multi-chain laundering dataset. 
We evaluate \textsc{FlowShield} on \textit{BybitML} and two public laundering datasets and 
experimental results demonstrate that \textsc{FlowShield} achieves the best overall performance, with an average F1 score of 98.0\%. 
Further behavior and SAR analyses demonstrate that \textsc{FlowShield} can reveal diverse laundering strategies and produce readable reports for investigating complex multi-hop fund flows.

\end{abstract}


\section{Introduction} \label{sec:intro}

Cryptocurrency money laundering (Crypto ML)~\cite{campbell2018bitcoin} causes substantial economic losses~\cite{26Crime}. Modern launderers fragment trails by chaining diverse intents such as direct transfers, token swaps, indirect transfers, and cross-chain transfers, dispersing stolen assets across many addresses and chains within minutes~\cite{report}. Meanwhile, decentralized finance enables transactions without centralized intermediaries~\cite{CASTILLOLEON2026102916}, weakening regulatory chokepoints and complicating timely intervention and attribution~\cite{ustreasury2023_defi_risk}. Consequently, effective cryptocurrency anti-money laundering (Crypto AML) is essential for risk control and regulatory review.

Existing Crypto AML methods~\cite{sok25fu} attempt to detect illicit transactions through heuristics, statistical features, or graph structures, but they do not jointly address semantic understanding, fund-flow tracking, and investigator-facing reporting. Rule-based heuristics rely on manually designed patterns~\cite{lv2023detection} or blacklists~\cite{wu2023towards}, making them fragile against evolving and composed laundering strategies~\cite{sok25fu}. Machine learning methods improve adaptability by learning from transaction features~\cite{elmougy2023demystifying,wang2024graphalm,zhou2023visual}, but often treat transactions as independent samples and struggle to reason over multi-hop fund flows. Graph-based methods model structural relations~\cite{alarab2020competence,lin2024denseflow,nicholls2023fraudlens,weber2019anti,wu2023tracer}, yet most remain address-centric or topology-driven, simplifying transaction semantics and lacking explicit fund-flow reconstruction. Moreover, their outputs are usually limited to detection labels rather than reports that summarize suspicious multi-hop flows for investigators.
To overcome the limitations of existing approaches and develop next-generation reliable Crypto AML solutions, we identify three core challenges as follows. 

% \begin{figure}[t]
% \centering
%   % \includegraphics[width=\linewidth]{Figs/example_ml_technique_2.png}
%     \includegraphics[width=0.7\linewidth]{Figs/bybit_case.pdf}
% \caption{A typical laundering segment of Bybit Hack.}
% \label{fig:intro}
% \end{figure}
%As fig show, different edges. (i.e., as figure~\ref{fig:intro} shows, edges with different colors )
% \textit{\textbf{Challenge 1: Transaction semantics understanding.}} 
% % Understanding transaction semantics is fundamental to accurately interpreting on-chain behaviors, 
% Sophisticated laundering pipelines often orchestrate diverse transaction intents to obfuscate fund provenance
% (e.g., the hybrid semantics in Figure~\ref{fig:intro}).
% However, these semantics are not explicitly encoded on-chain and must be inferred from transaction logs. 
% From a graph perspective, relying solely on topology is insufficient because structurally similar transactions can correspond to different intents (i.e., 0xf452 and Thorchain bridge in Figure~\ref{fig:intro} share a 1-in-1-out structure despite distinct intents), making it \textbf{challenging} to disambiguate laundering behaviors using structure alone.
% 比如：他们都是转账的形式，光从表面难以分辨之类的。
% 同一地址在两笔交易中都可能看似“收款”，但一笔是直接转账，另一笔却是在换币

\textit{\textbf{Challenge 1: Transaction semantics understanding.}}
Understanding transaction semantics is necessary for distinguishing heterogeneous laundering operations, since similar surface transfers can correspond to different behaviors such as swaps, bridge deposits, or contract-mediated forwarding.
For example, in Figure~\ref{fig:intro}, the transfer to 0xf452 and the Thorchain bridge deposit share a similar 1-in-1-out structure, but correspond to different transaction semantics.
However, such semantics are not explicitly encoded in raw on-chain fields and must be inferred from transfer groups and event logs.
Therefore, the challenge is to recover behavior-level semantics that are hidden behind similar surface transfer patterns.

%-----0206
%(i.e. As fig show, different edges)
% \textit{\textbf{Challenge 2: Fund flow tracking.}} Tracking fund flows is essential for understanding illicit value propagation, particularly when funds are split or merged. Nevertheless, in account-based blockchains, raw transaction data does not provide explicit links between incoming and outgoing transfers (e.g., distinguishing whether the 89 ETH outflow to 0xd242 stems from the 58 ETH transfer or the 200k DAI swap inflow at the OKX router in Figure~\ref{fig:intro}). Existing transaction graphs~\cite{lin2024denseflow, lv2023detection, wu2023towards,wu2023tracer} model transactions as discrete edges, neglecting the continuity of fund flows. Thus, tracking fund flow is \textbf{challenging} since implicit propagation paths should be inferred by transaction semantics and contextual information.
% 精简这句话 is challenging because
% When multiple downstream transfers exist, fund-flow propagation cannot be determined from structure alone and must instead be inferred by jointly considering transaction semantics and contextual information 
% %(e.g., time proximity and amount/token consistency), 
% making accurate flow tracking challenging.

\textit{\textbf{Challenge 2: Fund-flow tracking.}}
Tracking fund flows is essential for understanding how illicit value propagates across transactions, especially when funds are split or merged.
However, account-based blockchains do not explicitly link a specific inflow to a later outflow.
For example, in Figure~\ref{fig:intro}, it is unclear whether the 89 ETH outflow to 0xd242 originates from the 58 ETH transfer or from the 200k DAI swap inflow at the OKX router.
Existing transaction graphs~\cite{lin2024denseflow,lv2023detection,wu2023towards,wu2023tracer} model transfers as discrete edges, but do not capture how value continues across edges.
Therefore, the challenge is to infer local fund-propagation paths from transaction semantics and contextual signals.

\begin{figure}[t]
\centering
    \includegraphics[width=0.9\linewidth]{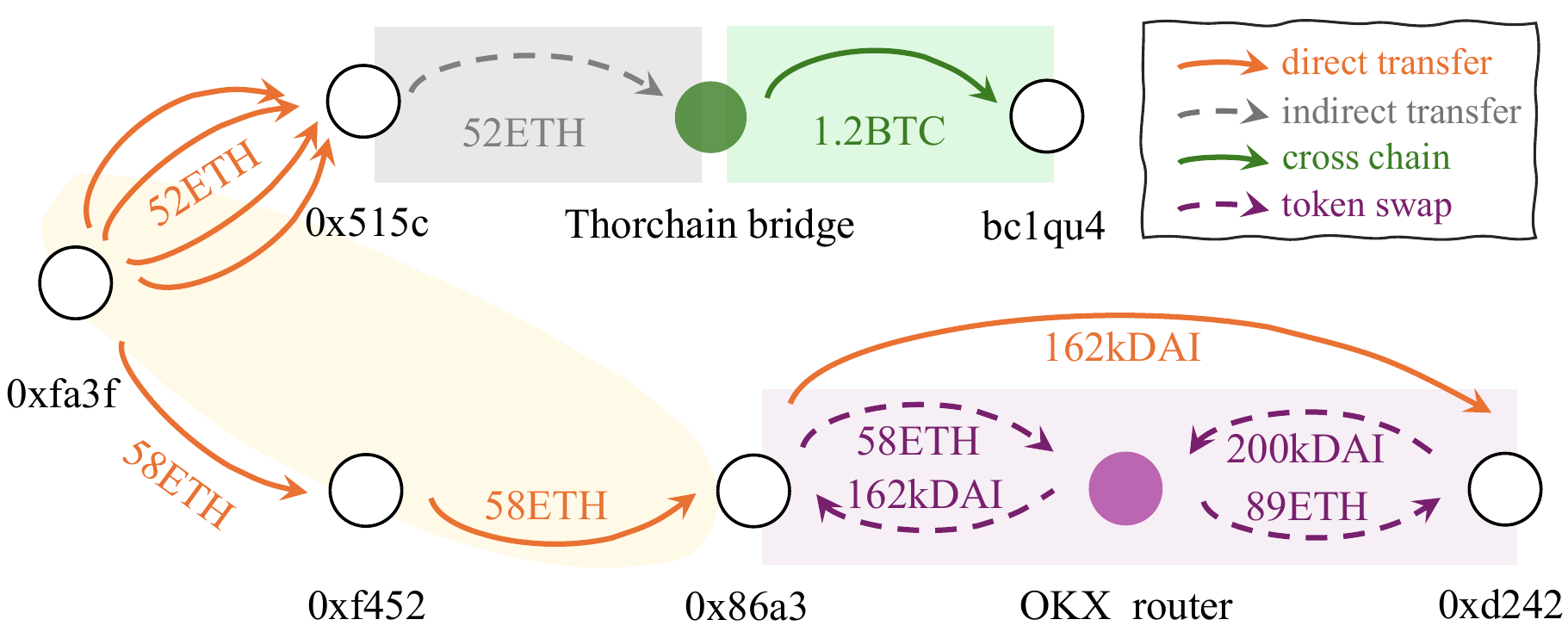}
% \caption{A typical laundering segment of Bybit Hack.}
\caption{A representative laundering segment from the Bybit Hack incident.}
\label{fig:intro}
\end{figure}

\textit{\textbf{Challenge 3: Heterogeneous multi-view learning.}}
Crypto AML signals are distributed across multiple incomplete but complementary views, including transaction semantics, fund-flow context, and graph structure.
Textual semantics and flow descriptions capture behavioral intent and local value movement, while graph structure captures topological dependencies among addresses.
Relying on any single view may miss important laundering cues.
Therefore, the challenge is to learn a unified laundering-relevant representation that integrates heterogeneous textual and structural signals for transaction-level AML detection.

\begin{table*}[t]
\centering
\scriptsize
% \qw{maybe more metrics?}
% \caption{Comparison of typical Crypto AML methods. $\text{S}_{dir}$, $\text{S}_{ind}$, $\text{S}_{swp}$, and $\text{S}_{crs}$ are four kinds of transaction semantics.}
\caption{Comparison of representative Crypto AML methods. $\text{S}_{dir}$, $\text{S}_{ind}$, $\text{S}_{swp}$, and $\text{S}_{crs}$ denote four kinds of transaction semantics.}
\label{tab:aml_comparison}

\resizebox{0.9\linewidth}{!}{
\begin{tabular}{l|c|c|c@{\hspace{1.6pt}}c|c@{\hspace{1.6pt}}c@{\hspace{1.6pt}}c@{\hspace{1.6pt}}c|c@{\hspace{1.6pt}}c}
\toprule
\multirow{2}{*}{\textbf{Crypto AML}} 
&  \multirow{2}{*}{\textbf{Method}} 
& \multirow{2}{*}{\textbf{Dataset (Year)}} 
& \multicolumn{2}{c}{\multirow{1}{*}{\textbf{Blockchain}}} 
& \multicolumn{4}{c}{\textbf{Semantics}} 
& \multirow{2}{*}{\makecell[c]{\textbf{Laundering}\\\textbf{flow}}}  
& \multirow{2}{*}{\makecell[c]{\textbf{Investigator}\\\textbf{report}}} \\
 \cmidrule(lr){6-9}
 % \textcolor{red!80}{P2}
 &  & & UTXO & Account & $\text{S}_{dir}$ & $\text{S}_{ind}$ & $\text{S}_{swp}$  & $\text{S}_{crs}$  &   \\
\midrule
Youssef\cite{elmougy2023demystifying} & Machine learning & \textit{Elliptic}~(2019) &\greencheck & \redx & \greencheck &  \redx & \redx &   \redx & \greencheck &  \redx \\
FraudLens\cite{nicholls2023fraudlens} & Graph-based       & \textit{Elliptic}~(2019)  & \greencheck & \redx & \greencheck &  \redx & \redx & \redx & \greencheck  & \redx \\
Weber\cite{weber2019anti} & Graph-based     & \textit{Elliptic}~(2019) & \greencheck & \redx & \greencheck &  \redx & \redx & \redx & \greencheck  & \redx \\
MPOCryptoML\cite{samadi2025mpocryptoml} & Graph-based & \textit{Elliptic}~(2019) & \greencheck & \redx & \greencheck & \redx & \redx & \redx & \greencheck & \redx \\
MRGNN\cite{hyun2023anti} & Graph-based & \textit{Not available} & \greencheck & \redx & \greencheck & \redx & \redx & \redx & \redx & \redx \\
Cyclic\cite{lv2023detection}    & Heuristic & \textit{Not available} & \redx & \greencheck &  \greencheck & \redx & \redx & \redx & \redx & \redx \\
XBlock\cite{wu2023towards}   & Heuristic & \textit{EthereumHeist}~(2022) &\redx & \greencheck &  \greencheck& \redx & \redx & \redx & \redx & \redx \\
DenseFlow\cite{lin2024denseflow} & Graph-based & \textit{EthereumHeist}~(2022) &\redx & \greencheck &  \greencheck& \redx & \redx & \redx & \redx & \redx \\
Tracer\cite{wu2023tracer}    & Graph-based & \textit{Not available} & \redx & \greencheck &  \greencheck&  \redx &  \greencheck & \redx &  \redx  & \redx \\
\midrule
\rowcolor{gray!10}
 \techName{}  & Graph + LLM    & \textit{BybitML}~(2025) &\greencheck & \greencheck &  \greencheck&  \greencheck&  \greencheck&  \greencheck&  \greencheck&  \greencheck\\
\bottomrule
\end{tabular}
}
\end{table*}

To address the above challenges, we present \textsc{FlowShield}, a new framework for detecting Crypto ML transactions and generating investigator-facing reports.
Specifically, to address \textit{\textbf{Challenge 1}}, \textsc{FlowShield} introduces a \emph{transaction semantics parsing module} that recovers behavior-level semantics from observable relations within transaction groups, identifying direct transfers, token swaps, indirect transfers, and cross-chain transfers.
To address \textit{\textbf{Challenge 2}}, \textsc{FlowShield} employs a \emph{fund-flow construction module} that reconstructs localized fund-flow subgraphs from upstream, downstream, and parallel perspectives, capturing potential funding sources, subsequent redistribution, and similar-amount parallel movements.
To address \textit{\textbf{Challenge 3}}, \textsc{FlowShield} further employs a \emph{text--structure fusion module} that encodes semantics and fund-flow texts with a large language model (LLM), encodes graph-structural context with a graph convolutional network (GCN), and fuses the two kinds of representations through a bidirectional interaction layer.
Finally, \textsc{FlowShield} uses a \emph{detection and report generation module} to identify laundering transactions and organize detected multi-hop fund flows into suspicious activity reports (SARs).

Our main contributions are summarized as follows:
% In summary, this paper makes the following key contributions:
\begin{packeditemize}
%C1: 新的洗钱行为建模方式。据我们所知，本文首次在加密货币反洗钱场景中同时考虑交易行为语义与资金传播过程进行建模，将洗钱检测从基于地址或交易的判别，扩展为对资金如何通过多种交易行为在交易网络中传播的刻画，从而更好地描述复杂洗钱过程。
    \item \textbf{A new modeling perspective.} To the best of our knowledge, this work is the first to jointly model \emph{transaction semantics} and \emph{fund-flow propagation} for Crypto AML. 
    We move beyond address- or transaction-level classification and instead model how funds propagate through diverse transaction behaviors, enabling a more faithful representation of complex laundering processes.

%C2: 创新的反洗钱检测框架。我们提出了一种新的反洗钱检测框架 \textsc{FlowShield}，由语义感知、资金流建模、语义–结构融合以及 SAR 生成四个模块组成，能够联合分析交易意图与资金流结构，实现对复杂洗钱行为的准确检测与可解释分析。
% qs:这里直接写via GCNs and LLM有点怪，不知道要怎么写合适
    \item \textbf{A unified detection and reporting framework.} We propose \textsc{FlowShield}, a novel Crypto AML framework consisting of four modules. 
    The framework jointly leverages textual behavioral descriptions and graph-structural context for transaction-level detection, and organizes detected multi-hop fund flows into investigator-facing SARs.
    
%C3: 首个公开的多链洗钱数据集。我们构建并公开了首个基于真实重大安全事件的多链洗钱数据集 \textit{BybitML}，覆盖 Ethereum 与 Bitcoin 两条主流区块链，并提供交易语义标注与洗钱交易标注，为相关研究提供可复现的实验基准。
    \item \textbf{The first multi-chain AML dataset.}  We construct and open-source BybitML\footnote{\textit{BybitML} is publicly available at \url{https://osf.io/nx7aj/overview?view_only=663038fbea43491bb4010011c7f28b23}.}, the first public multi-chain Crypto AML dataset derived from a real-world major security incident. 
    The dataset covers two major blockchains with transaction-level semantic annotations and laundering labels.
    
%C4：全面的实验验证。我们在 \textit{BybitML} 数据集上对 \textsc{FlowShield} 进行了系统评估，并与九种代表性基线方法进行了对比。实验结果表明，该方法在整体检测性能上显著优于现有方法。我们开源了数据集与代码。
    \item \textbf{Extensive experimental validation.} We conduct comprehensive evaluations of \textsc{FlowShield}\footnote{The implementation of \textsc{FlowShield} is available at \url{https://osf.io/9s3j2/overview?view_only=9b15fcf6978b47ddbdcc8a850d1666c9}.} on \textit{BybitML}  and two public laundering datasets, comparing it with thirteen representative baselines. 
    Experimental results demonstrate that \textsc{FlowShield} consistently achieves the best detection performance and remains robust across heterogeneous laundering scenarios.

% \item \textbf{Insights:} We perform an in-depth analysis of laundering transactions detected by \textsc{FlowShield}, summarizing diverse laundering strategies and evasion behaviors, which provides insights for regulatory and investigative practice.
    
\end{packeditemize}

\section{Related Work}

Existing Crypto AML techniques broadly fall into three categories.
% %Rule-based heuristics
% \textbf{\textit{Rule-based heuristics}} approaches rely on predefined rules derived from known laundering patterns, such as blacklists~\cite{wu2023towards}, cyclic flows~\cite{lv2023detection}, or mixer usage~\cite{kuccuk2023investigation}. These methods are straightforward for capturing known typologies, but fail against novel or obfuscated laundering strategies. 
% \textbf{\textit{Machine learning}} methods train classifiers on features extracted from transactions, leveraging techniques such as feature refinement~\cite{elmougy2023demystifying}, active learning~\cite{wang2024graphalm}, and anomaly detection~\cite{zhou2023visual}. 
% These methods are more adaptive but treat transactions independently, limiting multi-hop fund-flow reasoning.
% \textbf{\textit{Graph-based}} methods model the transactions as networks to uncover laundering accounts or transactions. Subgraph search (e.g., DenseFlow~\cite{lin2024denseflow}, Tracer~\cite{wu2023tracer}), graph embeddings (e.g., GTN2Vec~\cite{liu2023graph}, Sub2Vec~\cite{bellei2024shape}), and GNNs~\cite{alarab2020competence,lo2023inspection,nicholls2023fraudlens,hyun2023anti,guo2023lb,song2024illicit} have been applied. While capturing structure, most are address-centric and ignore transaction semantics and value propagation.
% Existing Crypto AML techniques broadly fall into three categories.
\textbf{\textit{Rule-based heuristic}} approaches rely on predefined rules derived from known laundering patterns, such as blacklists~\cite{wu2023towards}, cyclic flows~\cite{lv2023detection}, or mixer usage~\cite{kuccuk2023investigation}. These methods are effective for known typologies but brittle against evolving laundering strategies.
\textbf{\textit{Machine learning}} methods train classifiers on transaction or address features, including feature refinement~\cite{elmougy2023demystifying}, active learning~\cite{wang2024graphalm}, anomaly detection~\cite{zhou2023visual}, and CryptoML-style supervised classifiers. These methods improve adaptability but often treat transactions independently, limiting multi-hop fund-flow reasoning.
\textbf{\textit{Graph-based}} methods model transactions as graphs to detect suspicious addresses or transactions. 
GNN-based methods such as GCN~\cite{kipf2016semi} and GraphSAGE~\cite{hamilton2017inductive} learn representations from transaction graph topology, while MRGNN~\cite{hyun2023anti} further models multi-relational transaction graphs to capture heterogeneous interactions.
Graph pattern mining methods such as Tracer~\cite{wu2023tracer}, DenseFlow~\cite{lin2024denseflow}, and MPOCryptoML~\cite{samadi2025mpocryptoml} search subgraphs or laundering structures such as fan-in/out, bipartite, and gather-scatter patterns.
However, these methods either remain topology-driven or rely on predefined graph patterns without reconstructing transaction semantics and local fund-flow propagation.
% Table~\ref{tab:aml_comparison} summarizes representative methods and the main gaps in datasets, semantics, flow reasoning, and explainability. 
% Widely used public datasets (e.g., \textit{Elliptic}~\cite{weber2019anti} and \textit{EthereumHeist}~\cite{wu2023towards}) are single-chain and dated, with limited transaction details or heuristic-derived labels, limiting modern multi-chain evaluation. 
% prior methods abstract transactions as direct transfers, while only a few cover additional semantic such as token swap.
% On account-based blockchains, they do not explicitly trace value propagation. They rarely provide investigator-friendly explanations. So, we design \textsc{FlowShield} to fill these gaps by building a multi-chain dataset, supporting richer semantics, enabling explicit fund-flow reasoning on account-based chains, and generating explainable SARs.

Table~\ref{tab:aml_comparison} summarizes representative methods and the main gaps.
Widely used public datasets (e.g., \textit{Elliptic}~\cite{weber2019anti} and \textit{EthereumHeist}~\cite{wu2023towards}) are single-chain and dated, with limited transaction details or heuristic-derived labels, limiting modern multi-chain evaluation.
Prior methods mostly abstract transactions as direct transfers, while only a few cover additional semantics such as token swaps.
More importantly, even graph-based methods rarely explicitly trace how value propagates across transactions, especially on account-based blockchains where incoming and outgoing transfers are not directly linked.
They also rarely transform detection results into investigator-facing reports that summarize suspicious flows and red flags.
Motivated by these gaps, we design \textsc{FlowShield} to support multi-chain evaluation, richer transaction semantics, explicit fund-flow construction, and suspicious activity report generation.

\section{Problem definition}

\heading{Notations.}
Let $G=(\mathcal{V}, \mathcal{E}, \mathbf{X}, \mathbf{A})$ be a multi-directed transaction graph, where $\mathcal{V}=\{v_1,\ldots,v_n\}$ is the address set, $\mathcal{E}=\{t_1,\ldots,t_m\}$ is the transaction set, $\mathbf{X}\in\mathbb{R}^{n\times d_f}$ is the address feature matrix, and $\mathbf{A}\in\mathbb{R}^{n\times n}$ is the adjacency matrix with $A_{ij}$ denoting the number of transactions from $v_i$ to $v_j$.
Each raw transaction $t_k\in\mathcal{E}$ is represented as $(\mathrm{send}(t_k), \mathrm{recv}(t_k), \textit{h}_k, \textit{amount}_k, \textit{token}_k, \textit{time}_k, \textit{log}_k)$, where the fields denote sender, receiver, transaction hash, transferred value, asset type, timestamp, and transaction log, respectively.
The semantic type $\mathrm{S}_k\in\mathcal{S}$ is later assigned by the semantics parsing module (Sec.~\ref{subsec-semantic}). Since a transaction hash may correspond to multiple transfer records, including external, internal, and token transfers, we define its transaction group as $\mathcal{T}_{h}=\{t\in\mathcal{E}\mid \textit{hash}(t)=h\}$ for each hash $h\in\mathcal{H}$.
Figure~\ref{fig:exa} shows an example transaction group.

% \heading{Explainable Laundering Transaction Detection.}
\heading{Laundering Transaction Detection and Reporting.}
%给定多重有向交易图，我们的目标是对图中每一条边进行洗钱相关性检测，即学习一个边级分类器，为每条交易输出其为洗钱交易的预测标签。 在完成边级检测后，我们为每条可疑交易抽取其联通洗钱子图作为结构化证据。随后，将子图输入解释器，生成面向调查与复核的结构化可疑活动报告（SAR），其输出包含三部分：
% Given $G=\{\mathcal{V}, \mathcal{E}, \mathbf{X}, \mathbf{A}\}$, 
% our goal is to detect whether each transaction $t_k \in \mathcal{E}$ is money-laundering related. 
% Formally, we learn an edge-level classifier $f_{\Theta}$ that outputs a laundering label for each transaction:
% $\hat{y}_k = f_{\Theta}(G,t_k),\, \hat{y}_k \in \{0,1\}$,
% where $\hat{y}_k=1$ indicates a laundering transaction.
% After detection, we collect the flagged transfer set 
% $\hat{\mathcal{E}}_{ML}=\{t_k \in \mathcal{E}\mid \hat{y}_k=1\}$ and extract the connected subgraphs induced by 
% $\hat{\mathcal{E}}_{ML}$ in $G$, denoted as $\mathcal{G}_{ML}=\{G_1,\dots,G_M\}$.
% We then feed each $G_m \in \mathcal{G}_{ML}$ into an LLM-based report generator to produce an investigator-friendly suspicious activity report
% $r_m=\langle \textit{FlowDiagram}_m,\allowbreak \textit{Summary}_m,\allowbreak \textit{RedFlags}_m\rangle$,
% which contains a visualized flow diagram of $G_m$, a concise summary, and key red flags indicating suspicious behavior.
Given $G=(\mathcal{V}, \mathcal{E}, \mathbf{X}, \mathbf{A})$, our goal is to learn a transaction-level detection model~$f_{\Theta}$ that predicts a laundering label for each transaction:
$\hat{y}_k=f_{\Theta}(G,t_k)\in\{0,1\}$, where $\hat{y}_k=1$ indicates laundering.
After detection, we collect flagged transactions $\hat{\mathcal{E}}_{ML}=\{t_k\in\mathcal{E}\mid \hat{y}_k=1\}$ and extract their connected laundering flows $\mathcal{G}_{ML}=\{G_1,\ldots,G_M\}$.
For each detected flow $G_m$, \techName{} generates an investigator-friendly suspicious activity report~$r_m=\langle \textit{FlowDiagram}_m,\textit{Summary}_m,\textit{RedFlags}_m\rangle$.
\section{Methodology} \label{sec:approach}
\begin{figure*}[t]
\begin{center}
\includegraphics[width=0.92\textwidth]{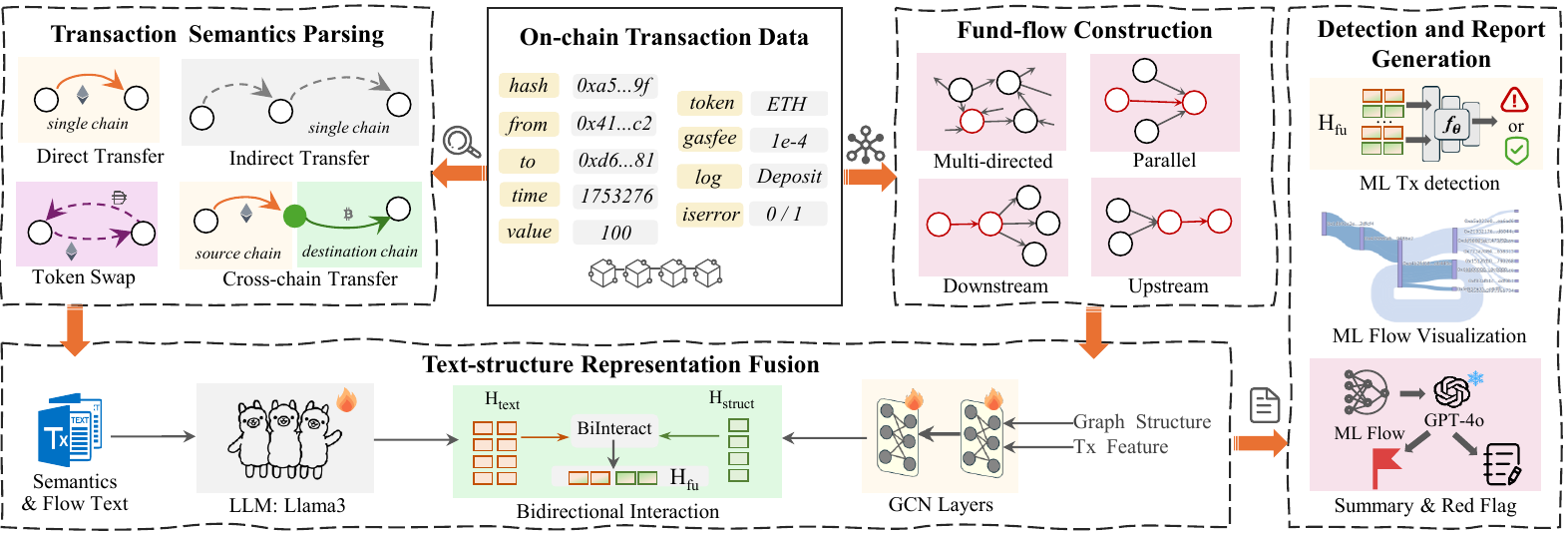}
\end{center}
% \caption{\label{fig:framework} An overall illustration of the proposed framework \techName{}.}
\caption{\label{fig:framework} Overview of the proposed \techName{} framework.}

\end{figure*}
% We first present an overview (\S\ref{subsec-archi}) of our framework. We then show the details of each phase (\S\ref{subsec-semantic}-\S\ref{subsec-sar}).

% \subsection{Architecture Overview}

\label{subsec-archi}

% In this section, we introduce \techName{} to improve the detection of complex money laundering activities while providing explainable outputs, with the pipeline demonstrated in Figure~\ref{fig:framework}. Firstly, to understand the intent behind a transaction, \textbf{ semantics parsing} module (Sec.~\ref{subsec-semantic}) provides transaction-level semantic understanding via pre-defined parsing rules based on the transaction logs and the relational structure. Next, to track value movement across transactions, \textbf{fund-flow construction} module (Sec.~\ref{fund_flow_aware}) builds three complementary subgraph views, tracing where transaction value originates, how it is redistributed downstream, and whether parallel flows occur. Then, to learn fused multi-view representations, \textbf{text-structure fusion} module (Sec.~\ref{subsec-fusion}) encodes semantics and flow texts with an LLM and structure with GCNs, and fuses the two views via cross-attention for detection. Finally, to provide explainable outputs, \textbf{SAR generation} module (Sec.~\ref{subsec-sar}) detects suspicious transactions with a lightweight classifier and prompts an LLM with the detected laundering flows to generate SARs with concise summaries and key red flags for investigation.
% \subsection{Overview of \techName{}}

In this section, we introduce \techName{} to improve the detection of complex money laundering activities while generating investigator-facing reports. The overall framework is illustrated in Figure~\ref{fig:framework}. 
Firstly, to understand the behavioral intent behind each transaction, we introduce a \textbf{semantics parsing} module (Sec.~\ref{subsec-semantic}), which extracts observable behavioral relationships from transaction groups to recover the semantic type.
Next, to track value movement across transactions, we introduce a \textbf{fund-flow construction} module (Sec.~\ref{fund_flow_aware}), which characterizes funding sources, downstream redistribution, and parallel transfer patterns while reducing irrelevant neighborhood noise. 
Then, to jointly leverage textual behavioral descriptions and graph-structural context, we introduce a \textbf{text--structure fusion} module (Sec.~\ref{subsec-fusion}). 
It jointly integrates textual behavior descriptions and graph-structural context into a unified transaction representation through a bidirectional interaction layer.
Ultimately, the fused representation is fed into a \textbf{detection and report generation} module (Sec.~\ref{subsec-sar}) to output transaction-level detection results and generate investigator-facing suspicious activity reports.

% 为了检测复杂的链上洗钱行为并生成面向调查人员的可读报告，我们提出 FlowShield，其整体框架如图所示。首先，为了理解单笔交易背后的行为意图，我们提出交易语义解析模块。该模块从转账组中提取可观测的行为关系，恢复交易的行为语义。其次，为了追踪交易之间的价值移动，我们提出资金流构建模块，能够在降低无关邻域噪声的同时，刻画资金来源、后续再分配和并行转移模式。随后，为了联合利用文本化行为描述与图结构上下文，我们提出文本-结构对齐模块，使模型能够在结构上下文中关注关键行为语义，并用语义-资金流描述补充纯图结构难以区分的行为差异。最后，为了输出交易级检测结果并生成面向调查人员的可读报告，我们提出检测与解释模块，先用分类器识别洗钱交易，随后基于检测到的多跳资金流生成可疑活动报告。

% \begin{packeditemize}

% \item Phase-\circlednum{1} (\emph{semantics parsing}) provides transaction-level semantic understanding by extracting high-level behavioral intents from transaction data. It can identify direct/indirect transfers, token swaps, and cross-chain transfers.  

% \item Phase-\circlednum{2} (\emph{fund-flow construction}) model transactions as a multi-directed graph and derives upstream, downstream, and parallel subgraphs to reveal fund flows. It overcomes the limitation of address-centric transaction graphs.

% \item Phase-\circlednum{3} (\emph{semantics–structure fusion}) integrates both views through fused embedding and yields a richer transaction representation. It bridges the gap between semantic and structural modeling.

% \item Phase-\circlednum{4} (\emph{SAR generation}) detects suspicious transactions and generates reports with flow visualisations, path summaries, and red flags. It provides explainable outputs.

% \end{packeditemize}

\subsection{Semantics Parsing}
\label{subsec-semantic}

Although transactions provide fields such as from/to/amount, the underlying intent behind a transaction is not directly exposed by these surface fields. 
For example, a single contract invocation may trigger multiple external transfers, internal transfers, and token movements, and the relations among these transfers reveal the actual behavioral intent.
If a model relies only on these fields or local topology, it may conflate transactions that are structurally similar but semantically different. 
Therefore, we 
% need to recover behavior-level semantics from transfer relations within transaction groups. 
% To achieve this, we 
introduce a semantics parsing module, which leverages observable signals to parse raw transactions into four transaction semantics.

\heading{Recovering transaction semantics.}~Given a transaction $t_k$, the semantics parsing module maps it to one of four transaction semantics:
$\mathcal{S}=\{\mathrm{S}_{dir}, \mathrm{S}_{swp}, \mathrm{S}_{ind}, \mathrm{S}_{crs}\}$,
corresponding to direct transfer, token swap, indirect transfer, and cross-chain transfer, respectively. 
We focus on these four semantics because they cover the core operations in laundering techniques.
% : direct transfers disperse funds across addresses,  token swaps break token-level continuity, indirect transfers hide sender--receiver relations, and cross-chain transfers move value beyond a single blockchain.
Except for direct transfer, the other three semantics are recovered from observable behavioral signals within the transaction group: token-flow consistency, intra-transaction forwarding, and bridge-log parsing.
If none of these signals is observed, the transaction is kept as a direct transfer.

% 为了恢复这些语义，我们检查与 $t_k$ 具有相同 transaction hash 的转账组 $\mathcal{T}_{h_k}$，并比较组内转账之间的发送方、接收方、代币类型和金额关系。
% To recover these semantics, we examine the transfer group $\mathcal{T}_{h_k}$ that shares the same transaction hash as $t_k$, and compare sender, receiver, token, and amount relations among transfers within the group.

\begin{packeditemize}

\item\underline{Token-flow consistency.}
A transaction $t_k$ is identified as a \textbf{token swap} when its sender sends out one token and receives another token within the same transfer group.
Formally, if there exists another transfer $t_\ell \in \mathcal{T}_{h_k}$ such that $\mathrm{recv}(t_\ell)=\mathrm{send}(t_k)$ and $\mathrm{token}_\ell \neq \mathrm{token}_k$, we assign $t_k$ to $\mathrm{S}_{swp}$.
Token swaps change the asset form of funds, breaking single-token trace continuity.

\item\underline{Intra-transaction forwarding.}
A transaction $t_k$ is identified as an  \textbf{indirect transfer} when its receiver further forwards the same token within the same transfer group.
Specifically, if there exists $t_\ell \in \mathcal{T}_{h_k}$ such that $\mathrm{send}(t_\ell)=\mathrm{recv}(t_k)$, $\mathrm{token}_\ell=\mathrm{token}_k$, and $\mathrm{amount}_\ell=\mathrm{amount}_k$, we assign $t_k$ to $\mathrm{S}_{ind}$.
Indirect transfers introduce intermediate addresses, obscuring the underlying sender--receiver relation.

\item\underline{Bridge-log parsing.}
A transaction $t_k$ is identified as a  \textbf{cross-chain transfer} when its receiver is a known bridge contract and the event log contains parseable destination-chain and recipient information.
Formally, if $\mathrm{recv}(t_k)\in\mathcal{A}_{bridge}$ and $\mathrm{parseable}(\mathrm{log}_k)$, we assign $t_k$ to $\mathrm{S}_{crs}$.
Cross-chain transfers move funds out of the current chain, making subsequent flows difficult to trace within a single-chain graph.

\end{packeditemize}

\begin{figure}[]
\begin{center}
\includegraphics[width=0.4\textwidth]{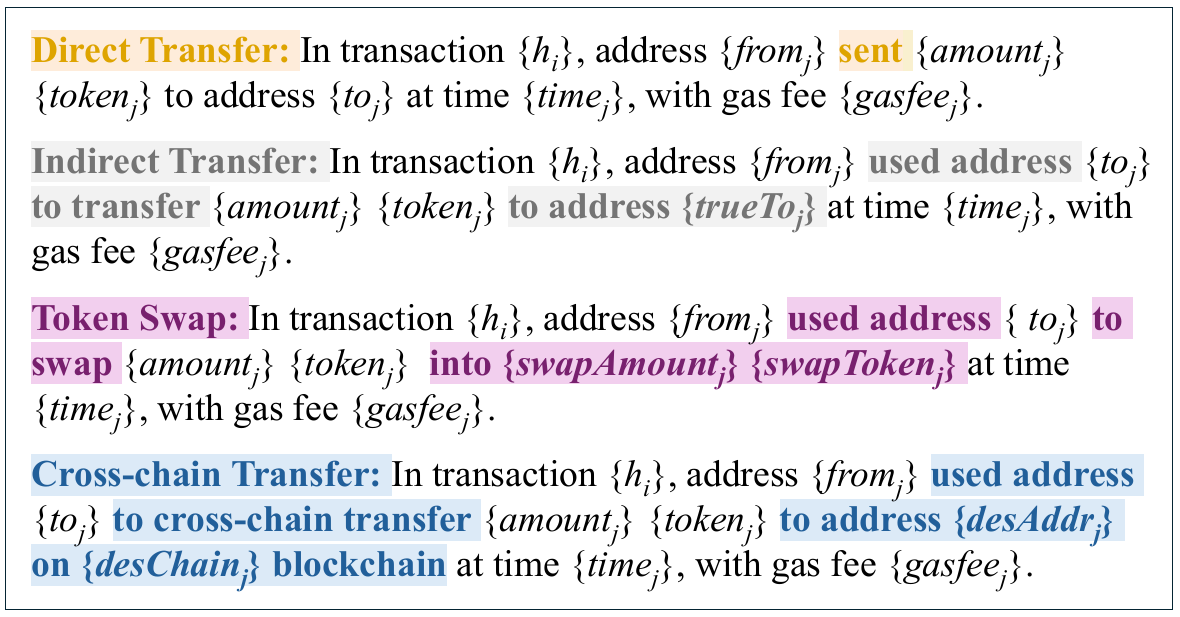}
\end{center}
% \caption{\label{fig:tmp} Semantics template text.}
\caption{\label{fig:tmp} Templates for generating semantic text.}
\end{figure}

\begin{figure}[]
\begin{center}
\includegraphics[width=0.32\textwidth]{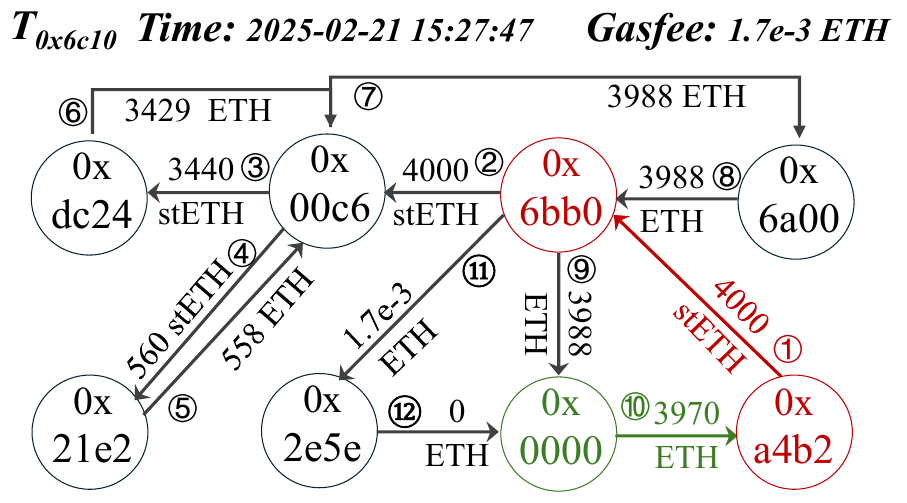}
\end{center}
\caption{\label{fig:exa}Example of token-swap parsing:  {0xa4b2} swaps stETH for ETH.}
\end{figure}

\heading{Generating semantic text.}
After recovering the semantic type, \techName{} instantiates a type-specific template with common transaction fields and semantic-specific fields, as shown in Figure~\ref{fig:tmp}. 
The semantics parsing turns raw transactions into intent-aware text, which supports downstream flow reasoning, reduces misclassification, and explainable reporting.

\heading{Running example.}
Figure~\ref{fig:exa} shows a token swap example. 
Within the same transfer group, address 0xa4b2 sends stETH and receives ETH, satisfying token-flow consistency and assigning $t_1$ to $\mathrm{S}_{swp}$.
\techName{} then instantiates the token-swap template to generate the semantic text.

\subsection{Fund-flow Construction}
\label{fund_flow_aware}

% 动机
% 仅有交易的语义信息仍不足以刻画真实洗钱过程，因为洗钱本质上是一个“资金传播”的动态链路，资金会在短时间内被汇集、拆分、混合。若直接纳入目标交易的完整邻域，将不可避免地引入大量良性交易噪声。
% 因此，我们构建交易的局部资金流图通过抽取互补的三类子图\emph{上游}（资金来源）、\emph{下游}（资金去向）与 \emph{并行}（相似资金流），并基于子图结构将资金传播归纳为典型模式并生成 flow text。
% Transaction semantics alone are still insufficient to characterize money laundering, because laundering is a dynamic process of fund propagation, where funds are rapidly aggregated, split, and mixed. If we directly include the full neighborhood of a target transaction, we would inevitably introduce substantial benign noise. Therefore, we construct localized fund-flow graphs by extracting three complementary subgraphs and further summarize propagation behaviors into typical patterns and a flow text.

% 交易语义 alone are still insufficient to 描述洗钱中的资金传播过程。洗钱交易通常不是孤立发生的，资金会在多笔交易之间被接收、转移、拆分或汇聚。若直接使用目标交易的完整邻域作为资金传播上下文，会引入大量与目标资金无关的交易噪声。为此，we construct localized fund-flow graphs by extracting three complementary subgraphs and further summarize propagation behaviors.
% 在构建这些子图时，由于 token swap 会改变资金的资产形态，我们使用语义解析后的 effective token 来保持资金流追踪中的代币一致性。
Transaction semantics alone are insufficient to capture laundering fund flows, where funds are often aggregated, split, and transferred across multiple transactions. Using the full neighborhood of a target transaction as its context would introduce substantial unrelated noise. To focus on relevant value movements, \techName{} constructs localized fund-flow graphs from three complementary perspectives: upstream subgraphs trace potential funding sources of the sender, downstream subgraphs capture subsequent redistribution by the receiver, and parallel subgraphs capture similar-amount movements within a close time window. Since token swaps change token types, \techName{} uses the effective token derived from semantics parsing to maintain token consistency during subgraph construction.

\begin{packeditemize}

\item \underline{{Upstream subgraph}} ($SubG_{up}$) captures transactions that fund the sender of a target transaction $t_{tg}$.
To build $SubG_{up}$, \techName{} searches within $\tau_t$ before $t_{tg}$, selects transactions to $t_{tg}$'s sender with the \emph{effective token}, and takes the earliest ones until their cumulative amount reaches $\tau_f$ of $t_{tg}$'s amount; together with $t_{tg}$, they form $SubG_{up}$.

% time window 
% $\tau_t$, flow ratio $\tau_f$, and amount similarity $\tau_a$

% \item \underline{\textit{Downstream subgraph}} ($SubG_{down}$) captures the outflow path of a target transaction $e_{tg}$.
% To construct $SubG_{down}$, \techName{} collects outgoing transactions from $e_{tg}$'s receiver within a look-ahead window $\tau_t$ that involve its \emph{effective token} (received for non-swap, obtained for swap). We keep the earliest transactions whose cumulative amount reaches $\tau_f$ of $e_{tg}$'s amount. Together with $e_{tg}$, they form $SubG_{down}$.

\item \underline{{Downstream subgraph}} ($SubG_{down}$) captures the outflows following a target transaction $t_{tg}$.
To build $SubG_{down}$, \techName{} selects outgoing transactions from  $t_{tg}$'s receiver within $\tau_t$ after $t_{tg}$ that involve the \emph{effective token}, and keeps the earliest ones whose cumulative amount reaches $\tau_f$ of $t_{tg}$'s amount. Together with $t_{tg}$, they form $SubG_{down}$.

% \item \underline{\textit{Parallel subgraph}} ($SubG_{para}$) captures flows occurring in parallel with a target transaction $e_{tg}$. 
% To construct $SubG_{para}$, \techName{} collects incoming transactions to the receiver of $e_{tg}$ within $\tau_t$ after $e_{tg}$ that involve the \emph{effective token} of $e_{tg}$. 
% We then keep those whose amounts are within the similarity threshold $\tau_a$ of $e_{tg}$'s amount. Finally, these parallel transactions, together with $e_{tg}$,  constitute  $SubG_{para}$.

\item \underline{{Parallel subgraph}} ($SubG_{para}$) captures concurrent flows around a target transaction $t_{tg}$.
To build $SubG_{para}$, \techName{} selects incoming transactions to $t_{tg}$'s receiver within $\tau_t$ after $t_{tg}$ that involve the \emph{effective token}, and keeps those with amounts within $\tau_a$ of $t_{tg}$'s amount. Together with $t_{tg}$, they form $SubG_{para}$.

\end{packeditemize}

\techName{} further summarizes the target transaction's fund movement into five patterns based on its subgraphs: single outflow, entire disperse, partial disperse, mix, and no forward. Details appear in Algorithm~\ref{alg:flow_awareness}. It then generates a flow text summarizing fund sources and outflows, including distribution and temporal characteristics, using the template in Figure~\ref{fig:tmp_flow_}. This provides a compact yet informative summary of fund-flow behavior for downstream detection.

\begin{figure}[]
\begin{center}
\includegraphics[width=0.41\textwidth]{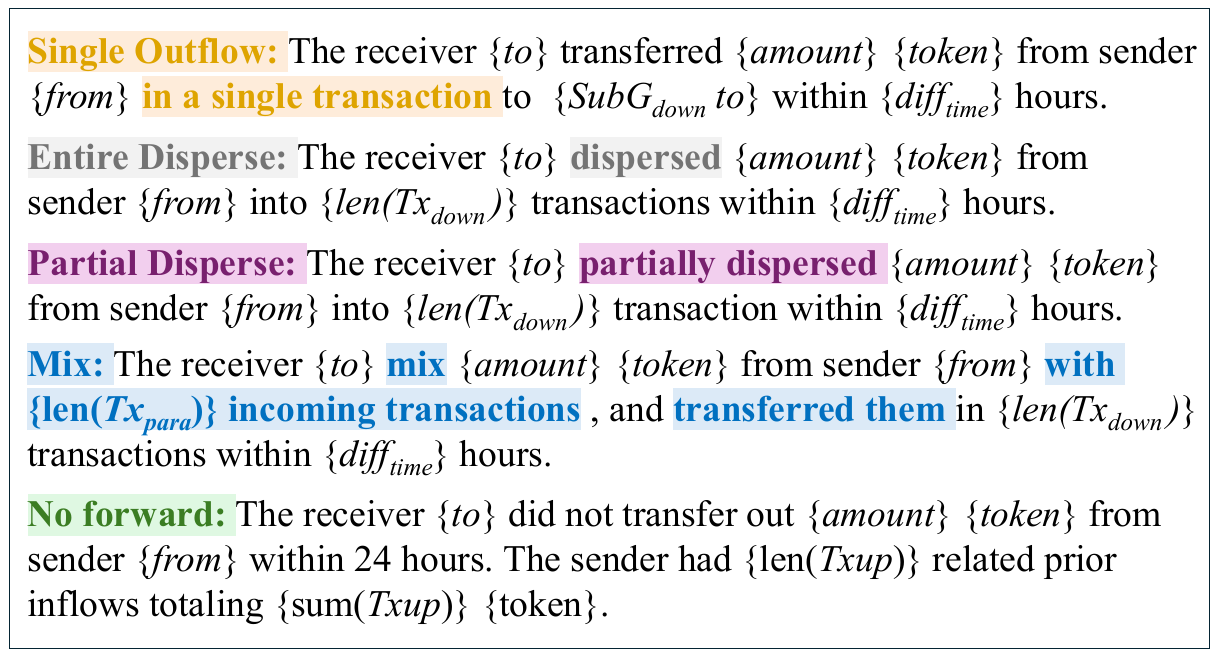}
\end{center}
% \caption{\label{fig:tmp_flow_} Flow template text. }
\caption{\label{fig:tmp_flow_} Template for generating fund-flow text.}
\end{figure}

\begin{figure}[]
\begin{center}
\includegraphics[width=0.34\textwidth]{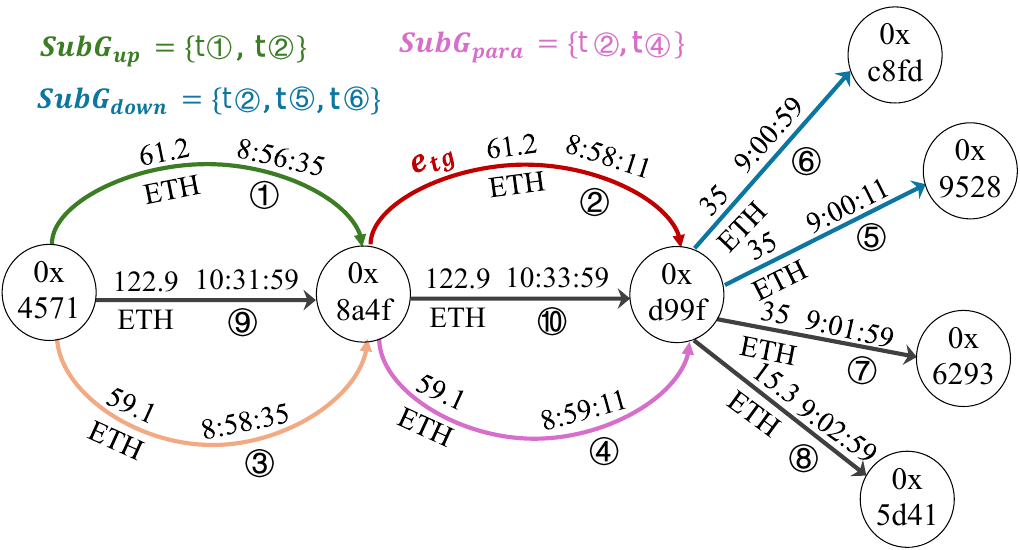}
\end{center}
\caption{\label{fig:exa_g} Example of fund-flow construction for target transaction $t_{2}$.}
\end{figure}

\heading{Running example.}
Figure~\ref{fig:exa_g} illustrates  $SubG_{up}$, $SubG_{down}$, and $SubG_{para}$ for the target transaction \hlhref{https://etherscan.io/tx/0xb570cf9587c99e978038c85e6f05d2fc50d12031ba3809102c66f71343b306c8}{$t_{2}$} under the thresholds in Sec.~\ref{sec:exps}.
\hlhref{https://etherscan.io/tx/0xe0a79310d14952ece3f7e72fae8c82aaf4c5d9d766918526a6a70b858f4e64ce}{$t_{1}$} funds $t_{2}$ and constitutes $SubG_{up}$.
The earliest outflows \hlhref{https://etherscan.io/tx/0x12a9f9347630615465c91d8bd79a21524ef2b73ff7a2ff2c3b3feeda112422d2}{$t_{5}$} and \hlhref{https://etherscan.io/tx/0x26540007067dc02e18948ff8e8038c552b7dffc81dce451a0d659b7d8bbc6dcb}{$t_{6}$} cumulatively reach the flow-ratio threshold $\tau_f$, forming $SubG_{down}$.
\hlhref{https://etherscan.io/tx/0x34bd630718364bec3859dc053f0ba202a85ae30f13c8365b47b647e84244a855}{$t_{4}$} is a similar-scale inflow, forming $SubG_{para}$.
As $t_{2}$ disperses over $\tau_f$ amount via multiple outflows, the pattern is \textit{entire disperse}.

% \heading{Running example.}
% Figure~\ref{fig:exa_g} illustrates $SubG_{up}$, $SubG_{down}$, and $SubG_{para}$ for the target transaction \hlhref{...}{$t_{2}$} under the thresholds in Sec.~\ref{sec:exps}.
% \hlhref{...}{$t_{1}$} is selected as a related prior inflow and forms $SubG_{up}$.
% The subsequent outflows \hlhref{...}{$t_{5}$} and \hlhref{...}{$t_{6}$} cumulatively reach the flow-ratio threshold $\tau_f$, forming $SubG_{down}$.
% \hlhref{...}{$t_{4}$} is a similar-scale incoming transfer to the receiver of $t_2$ within the amount-similarity threshold $\tau_a$, forming $SubG_{para}$.
% Since $t_{2}$ disperses more than $\tau_f$ of its amount via multiple downstream outflows, \techName{} summarizes its fund-flow pattern as \textit{entire disperse} and instantiates the flow template in Figure~\ref{fig:tmp_flow_}.

% % \vspace{-0.1in}
% \begin{center}
%   % \fbox{
% \colorbox{blue!9}{
% \begin{minipage}{0.95\linewidth}
%   (\textit{Entire disperse}) Receiver \hlhref{https://etherscan.io/address/0xd99f81773bb3b7eED44BD21C297Fd352Bccc6C5d}{0xd99f} dispersed 61.2 ETH from sender \hlhref{https://etherscan.io/address/0x8a4f03ca00400795119934c2ed458d8dd74a1dad}{0x8a4f} into 2 transactions within 2 minutes.
% \end{minipage}
%       }
%  %  }
% \end{center}

\subsection{Text-structure Fusion}
% \subsection{Text-structure Alignment}
\label{subsec-fusion}

% 语义文本和资金流文本显式描述交易意图与局部资金传播，但它们无法刻画交易在整体交易图中的结构位置。相反，GCN 能编码地址之间的拓扑上下文，但难以区分结构相似而行为语义不同的交易。为此，FlowShield 分别编码文本表示和结构表示，并通过双向 cross-attention 建模二者之间的交互，从而得到用于检测的统一交易表示。
Semantics and fund-flow texts explicitly describe transaction intent and local fund propagation, but they do not capture the transaction's structural position in the global transaction graph. In contrast, GCN-based structural embeddings encode topological context among addresses, but may conflate transactions with similar graph structures yet different behavioral meanings. To bridge these two views, \techName{} separately encodes textual and structural information into transaction-level representations, and then combines them through a bidirectional interaction layer for  detection.

\begin{packeditemize}

\item\underline{{Text embedding via LLM}.}
% \item As Figure~\ref{fig:llm} shows, 
\techName{} concatenates the semantic text and flow text as $\text{Txt}=\mathrm{Concat}(\text{Txt}_{sem}, \text{Txt}_{flow})$.
To adapt Llama3~\cite{grattafiori2024llama} to on-chain transaction semantics and fund-flow descriptions, we perform supervised instruction tuning~\cite{wei2021finetuned} with LoRA~\cite{hu2022lora}. Each training instance uses only $\mathrm{Txt}$ as the textual input, while the ML/normal label is used only as the supervised target for LoRA optimization and is not included in $\mathrm{Txt}$.
After tokenization, the LoRA-tuned Llama3 encoder produces final-layer hidden states $\mathbf{U}=[\mathbf{u}_1,\ldots,\mathbf{u}_m]\in\mathbb{R}^{m\times d_t}$.

We apply attention-mask mean pooling to obtain the transaction-level text representation:
%\[
\begin{equation}
\mathbf{h}_{text}
=
\frac{\sum_{i=1}^{m} a_i \mathbf{u}_i}{\sum_{i=1}^{m} a_i}
\in \mathbb{R}^{d_t},
\label{eq:text-pooling}
\end{equation}
%\]
where $\mathbf{u}_i$ is the final-layer hidden state of the $i$-th token, $a_i$ is its attention-mask value, $m$ is the number of text tokens, and $d_t$ is the Llama hidden dimension.

% \begin{figure}[t]
% \begin{center}
% % \includegraphics[width=0.48\textwidth]{Figs/text_embed_big.png}
% \includegraphics[width=0.4\textwidth]{Figs/lora.pdf}
% \end{center}
% \caption{\label{fig:llm} Encode text embedding via fine-tuning LLM. }
% \end{figure}

\item\underline{{Structure embedding via GCN.}}
\techName{} encodes graph-structural context by applying a two-layer GCN to the transaction MultiDiGraph.
Node embeddings are updated as
% \[
\begin{equation}
\mathbf{H}^{(l+1)}
=
\sigma\!\left(
\tilde{\mathbf{D}}^{-\frac{1}{2}}
\tilde{\mathbf{A}}
\tilde{\mathbf{D}}^{-\frac{1}{2}}
\mathbf{H}^{(l)}
\mathbf{W}^{(l)}
\right),
\label{eq:struct-embed}
\end{equation}
% \]
where $\tilde{\mathbf{A}}$ is the adjacency matrix with self-loops and $\tilde{\mathbf{D}}$ is the corresponding degree matrix. For a transaction, we form its structural representation by concatenating the final-layer GCN embeddings of its sender and receiver:
% \[
\begin{equation}
\mathbf{h}_{struct}
=
[\mathbf{z}_{src};\mathbf{z}_{dst}]
=
[\mathbf{H}^{(L)}_{src};\mathbf{H}^{(L)}_{dst}]
\in \mathbb{R}^{2d_g},
\label{eq:struct-con}
\end{equation}
% \]
where $d_g$ is the GCN hidden dimension and $L$ is 2.

\item\underline{{Bidirectional interaction.}}
To combine textual behavioral descriptions with transaction graph context, \techName{} uses a bidirectional interaction layer over transaction-level representations.
We first project the text and structural representations into a shared latent space:
$\tilde{\mathbf{h}}_{text}=\mathbf{W}_{text}\mathbf{h}_{text}$ and
$\tilde{\mathbf{h}}_{struct}=\mathbf{W}_{struct}\mathbf{h}_{struct}$.
The interaction layer transforms the two projected representations in both directions:
\begin{equation}
(\mathbf{f}_{s2t}, \mathbf{f}_{t2s})
=
\mathrm{BiInteract}(\tilde{\mathbf{h}}_{struct}, \tilde{\mathbf{h}}_{text}),
\end{equation}
where $\mathrm{BiInteract}(\cdot)$ denotes a bidirectional attention-based interaction layer.
The two directional outputs are concatenated and normalized to obtain the fused transaction representation:~$\mathbf{h}_{fused}
=
\mathrm{LayerNorm}
\left(
[
\mathbf{f}_{s2t}
\|
\mathbf{f}_{t2s}
]
\right).$
The fused representation is used for detection.

\end{packeditemize}

% \subsection{SAR Generation} 

\subsection{Detection and Report Generation} 
\label{subsec-sar}

% 设计动机
% AML 不仅要检测可疑交易，还要提供可复核的证据与可读解释；仅有分类标签难以支持调查与合规。 
% 因此，我们将检测结果结构化为 SAR：MLP 标记可疑交易，重建并可视化多跳资金流，再结合语义与流证据提示 LLM 生成 Summary 与 RedFlags，提升可解释性与可审计性。

% Crypto AML requires not only detecting suspicious transactions, but also providing human-readable explanations. A binary label alone is insufficient for investigation. Therefore, we structure the detection outputs into SARs: we first flag suspicious transactions with an MLP, and then prompt an LLM with semantic and flow evidence to generate a concise \textit{Summary} and key \textit{RedFlags}, improving explainability and auditability. 

% Crypto AML requires not only detecting suspicious transactions, but also providing human-readable explanations. 
% A binary label alone is insufficient for investigation. 
% Therefore, we first flags suspicious transactions with an MLP, then prompt an LLM to structure the outputs into SARs consisting of a \textit{FlowDiagram}, a concise \textit{Summary}, and key \textit{RedFlags}, improving explainability and auditability.

Crypto AML requires not only transaction-level detection, but also readable reports that help investigators inspect suspicious fund flows. A binary label alone does not show how flagged transactions are connected across multi-hop movements. Therefore, \techName{} first detects laundering transactions with an MLP classifier and then organizes the detected fund-flow subgraphs into investigator-facing SARs.

% In Phase-\circlednum{4}, \techName{} compiles detection results into investigator-friendly SARs. 
% Specifically, it (i) flags suspicious transactions, (ii) constructs a \textit{FlowDiagram} by visualizing multi-hop flows, and (iii) prompts GPT-4o to produce the \textit{Summary} and \textit{RedFlags}.

\begin{packeditemize}

\item\underline{{ML transaction detection.}} 
Given the fused transaction representation $\mathbf{h}_{fused}$, \techName{} uses an MLP classifier~\cite{tolstikhin2021mlp} to predict whether a transaction is money laundering:
% \[
\begin{equation}
\hat{y} = \arg\max_{k \in \{0,1\}}
\left[ W_2 \cdot \mathrm{ReLU}(W_1 h_{fused} + b_1) + b_2 \right]_k,
% \]
\end{equation}
where $W_1, W_2$ and $b_1, b_2$ are trainable parameters, $\mathrm{ReLU}(\cdot)$ is the rectified linear unit, and $[\cdot]_k$ denotes the $k$-th logit. 
$\hat{y}=1$ indicates money laundering, and $\hat{y}=0$ indicates normal behavior.

% \item\underline{\textit{ML flow visualization.}} 
% For transactions flagged as money laundering, the visualization module reconstructs the associated fund-flow subgraphs and merges related flows into a Sankey diagram, providing a holistic view of multi-hop asset movements.

\item\underline{{FlowDiagram visualization.}} 
% For transactions flagged as money laundering, the visualization module reconstructs the associated fund-flow subgraphs and merges related flows into a \textit{FlowDiagram} (Sankey diagram), providing a holistic view of multi-hop asset movements.
For transactions flagged as money laundering, \techName{} retrieves their associated multi-hop fund-flow subgraphs and merges related transactions into a \textit{FlowDiagram} (Sankey diagram), providing a visual summary of asset movements.

% \item\underline{\textit{SAR summary and red flags.}} 
% For each detected laundering flow, \techName{} aggregates the transaction descriptions and the flow evidence into a structured prompt, requesting (i) a concise flow summary and (ii) a list of red flags that indicate suspicious behavior. 
% We use GPT-4o for its strong long-context understanding and reasoning capabilities~\cite{shahriar2024putting}.

\item\underline{{SAR summary and red flags.}} 
For each flagged flow, \techName{} aggregates the transaction descriptions into a structured prompt, asking the LLM to summarize the laundering process and extract key red flags. 
We use GPT-4o for its strong long-context understanding and reasoning capabilities~\cite{shahriar2024putting}.
The generated SARs are intended to support manual review by summarizing the detected flow and highlighting suspicious behavioral cues.

\end{packeditemize}

\input{chapter/Alg-flow}

\section{Experimental Setup} \label{sec:exp-set}

% detailed descriptions of how data was collected, preprocessed, managed, and analyzed are required in AI4Sci track. 
% \subsection{Data Description}

% \subsection{Experiment Setup}

% \begin{figure}[t]
% \begin{center}
% \includegraphics[width=0.38\textwidth]{Figs/daily_stats.pdf}
% \end{center}
% \caption{\label{fig:sum_daily} Daily distribution of transactions in \textit{BybitML}. }
% \end{figure}

\begin{figure}[t]
  \centering
  \includegraphics[width=0.37\textwidth]{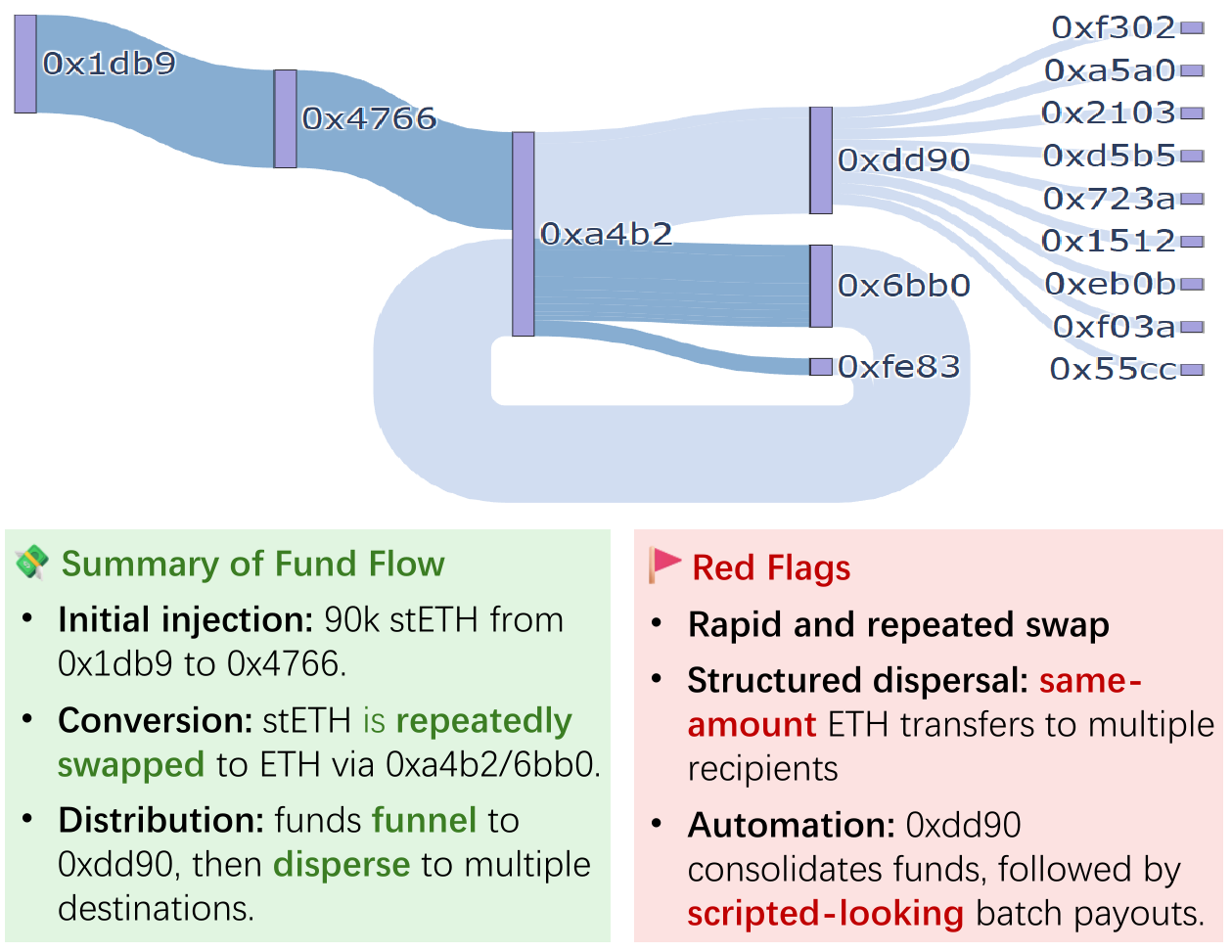}
  % \caption{\label{fig:sar_example}Example SAR generated by \techName{}, including a \textit{FlowDiagram}, \textit{Summary}, and \textit{RedFlags}.}
  \caption{\label{fig:sar_example}Abridged SAR example generated by \techName{}.}

\end{figure}

\input{Tabs/tab_effectiveness_avg_new}

\input{Tabs/tab_comparative_exp_new}

% The stacked bars represent the number of normal and  ML transactions, and the purple line indicates the  ML ratio. 

% The two commonly used public Crypto AML datasets, \textit{Elliptic}~\cite{weber2019anti} and \textit{EthereumHeist}~\cite{wu2023towards}, are both single-chain and relatively old, making them insufficient to capture the increasingly complex laundering behaviors and unsuitable for evaluation in multi-chain settings. Specifically, \textit{Elliptic} mainly offers highly normalized numerical features without essential transaction details (e.g., amount and time), thus limiting behavior-level modeling. \textit{EthereumHeist} derives laundering labels largely from heuristic rules rather than authoritative ground truth, which may introduce label noise.
% \heading{\textbf{Dataset.}}
\subsection{Dataset}
%https://www.elliptic.co/blog/bybit-exploit-six-months-on
We evaluate \techName{} on four real-world laundering datasets collected from three real laundering incidents. 
Among them, \textit{Bybit-EC} and \textit{Bybit-BC} form our contributed \textit{BybitML} dataset, the first public multi-chain dataset built from the 2025 Bybit Hack~\cite{bybit25}, one of the largest real-world laundering incidents to date.
We use blockchain APIs~\cite{tokenview_blockchain_api,blockstream_explorer_api} to collect Ethereum and Bitcoin transactions in the two-week high-frequency laundering window~\cite{elliptic_bybit_six_months},
% ~assigning transaction labels using intelligence sources: laundering transactions are labeled according to official Bybit laundering tags from blockchain security companies~\cite{elliptic_data_fabric}, and normal transactions are identified using the normal tag from \hlhref{https://intel.arkm.com/}{Arkham}~\cite{arkham_intel_platform}. 
~assigning transaction labels using intelligence sources: laundering transactions are labeled according to official Bybit laundering tags from blockchain security companies~\cite{elliptic_data_fabric}, and normal transactions are identified using the normal tag from \hlhref{https://intel.arkm.com/}{Arkham}~\cite{arkham_intel_platform}.  To assess label reliability, we manually audited 800 randomly transactions, covering 200 laundering and 200 normal transactions from each chain, and all audited labels were consistent with their assigned labels.
This results in two synchronized subsets, \textit{Bybit-EC} and \textit{Bybit-BC}, supporting realistic multi-chain evaluation.
Figure~\ref{fig:sum_daily} shows the daily distribution of transactions in \textit{BybitML}.
We further include \textit{Upbit} and \textit{AscendEX}, two public laundering datasets collected by XBlock~\cite{wu2023towards} through heuristic search from two real-world incidents. Together, these four datasets cover both account-based and UTXO-based blockchains, as well as different laundering scales and label distributions. Table~\ref{tab:dataset_stats} summarizes the statistics of all datasets.

%See Appendix~\ref{apd-data} for more details.

\input{Tabs/tab_dataset_stats}

% \subsection{Experiment Setup}
\subsection{Baselines and Metric}
We compare \textsc{FlowShield} with the state-of-the-art methods. According to the model architecture, we categorize the thirteen methods into four groups: 
\begin{packeditemize}
  % \item \textit{heuristic}: 
  % Cyclic~\cite{lv2023detection} detects cyclic transaction patterns, and XBlock~\cite{wu2023towards} uses blacklist poisoning to identify transactions linked to illicit addresses.
  \item \textit{Heuristic}: Cyclic~\cite{lv2023detection} detects cyclic patterns, and XBlock~\cite{wu2023towards} uses blacklist poisoning to flag laundering transactions.
    
  % \item \textit{machine learning}: 
  % Multi-Layer Perceptron (MLP)~\cite{rosenblatt1958perceptron}, Logistic Regression (LR)~\cite{cox1958regression}, and Support Vector Machine (SVM)~\cite{cortes1995support} are trained on  statistical features for ML detection~\cite{elmougy2023demystifying}.
  \item \textit{Machine learning}: MLP~\cite{rosenblatt1958perceptron}, LR~\cite{cox1958regression}, and SVM~\cite{cortes1995support} are trained on statistical features for ML detection~\cite{elmougy2023demystifying}.
  % \item \textit{graph learning}: 
  % GCN~\cite{kipf2016semi} and GraphSAGE~\cite{hamilton2017inductive} leverage transaction graph structure for ML detection~\cite{Mara24anti}, while Tracer~\cite{wu2023tracer} tracks fund flows through subgraph search.
  \item \textit{GNN-based}: GCN~\cite{kipf2016semi} and GraphSAGE~\cite{hamilton2017inductive} learn node representations from transaction graph structure for ML detection, while MRGNN~\cite{hyun2023anti} models multi-relational transaction graphs to capture heterogeneous interactions.

  % \item \textit{graph learning}: GCN~\cite{kipf2016semi} and GraphSAGE~\cite{hamilton2017inductive} leverage graph structure~\cite{Mara24anti}, while Tracer~\cite{wu2023tracer} tracks fund flows via subgraph search.

  \item \textit{Graph pattern mining}: Tracer~\cite{wu2023tracer} and DenseFlow~\cite{lin2024denseflow} search transaction subgraphs or fund-flow patterns, and MPOCryptoML~\cite{samadi2025mpocryptoml} detects multiple laundering structures such as fan-in/out, bipartite, and gather-scatter.

\item \textit{Multimodal graph}: GraphGPS~\cite{rampavsek2022recipe} and GraphFormers~\cite{yang2021graphformers} combine graph structure with Llama3-encoded transaction texts to evaluate whether multimodal graph models can benefit from semantic signals.

\end{packeditemize}

Given that the task focuses on detecting laundering transactions, which are regarded as positive samples, we use F1 score as the primary metric, as it jointly reflects precision and recall in imbalanced classification. Each experiment is repeated five times and we report the mean and standard deviation. 

% \heading{Metric.}  
% % We evaluate detection performance using four mainstream metrics:  {Precision (P)},  {Recall (R)},  {F1 score (F1)}, and  {money coverage rate (MCR)} (definition deferred to Appendix~\ref{apd-metrics}).
% % We report Precision (P), Recall (R), {F1 score (F1)}, and money coverage rate (MCR) (definition deferred to Appendix~\ref{apd-metrics}).
% We report F1 score (F1) as the primary metric. Each experiment is repeated five times, and we report the mean and standard deviation.

% \subsection{Implementation Details}
% All experiments were performed on Python 3.13.12, 1 NVIDIA RTX~4090 (24GB) with 46GB RAM.
% For fund–flow construction, we set the time window~$\tau_t=24$, the flow ratio~$\tau_f=0.9$, and the amount similarity~$\tau_a=0.05$. 
% The text encoder is Llama3-8B fine-tuned with LoRA using 5,000 transactions sampled only from the training split, and transaction-level embeddings are obtained by mean pooling the final hidden states.
% The structural encoder is a two-layer GCN with hidden dimension 16, producing 32-dimensional edge embeddings. The text and structural embeddings are fused using bidirectional cross-attention with 4 heads and dropout 0.1, followed by a two-layer MLP classifier with hidden dimension 64. The classifier is trained with Adam using learning rate 1e-3, batch size 64, and 100 epochs.

\subsection{Implementation Details}
All experiments were run on Python 3.13.12, 1 NVIDIA RTX~4090 with 46GB RAM. We use a 6:2:2 train/validation/test split for each dataset and optimize the classifier with cross-entropy loss.
For fund-flow construction, we set the time window~$\tau_t=24$, the flow ratio~$\tau_f=0.9$, and the amount similarity~$\tau_a=0.05$.
The text encoder is Llama3-8B fine-tuned with LoRA using 5,000 transactions sampled only from the training split of each run, and transaction-level embeddings are obtained by mean pooling the final hidden states.
The structural encoder is a two-layer GCN with hidden dimension 16, producing 32-dimensional transaction-level structural embeddings by concatenating the sender and receiver node embeddings.
The text and structural embeddings are projected into a shared latent space and fused using the bidirectional interaction layer with 4 attention heads and dropout 0.1, followed by a two-layer MLP classifier with hidden dimension 64.
The classifier is trained with Adam using learning rate 1e-3, batch size 64, and 100 epochs.

% \heading{experiment equipment.}
% All experiments were conducted on a workstation equipped with an NVIDIA RTX~4090 GPU (24GB VRAM) and 46 GB system memory.
% \heading{Hardware.} NVIDIA RTX~4090 (24GB) with 46GB RAM.

\section{Experimental Results} \label{sec:exps}

In this section, we analyze our experimental results to demonstrate the efficacy of \techName{}. Particularly, we aim to address the following research questions:
% \begin{packeditemize}
%     \item \textbf{RQ1:} Does \techName{} outperform various baselines?
%     \item \textbf{RQ2:} How does each component of \techName{} contribute to the performance enhancement?
%     \item \textbf{RQ3:} How sensitive is \techName{} to different hyperparameters?
%     \item 

%   \end{packeditemize}
\begin{packeditemize}
    \item \textbf{RQ1:} Does \techName{} outperform various baselines?
    \item \textbf{RQ2:} How does each component of \techName{} contribute to the performance enhancement?
    \item \textbf{RQ3:} Is \techName{} robust to hyperparameter changes and semantic parsing drift?
\item \textbf{RQ4}: Can \techName{} scale to large transaction volumes with low computational overhead?

\end{packeditemize}

\subsection{Performance Comparisons~(RQ1)}

We compare \techName{} with representative baselines and report both overall F1 scores across four datasets and day-by-day F1 scores on \textit{Bybit-EC}. As shown in Table~\ref{tab:avg_metrics_new},~\ding{182}~\textbf{\techName{} achieves the highest F1 on all datasets}.
Although the four datasets differ in blockchain type, scale, laundering ratio, and laundering strategies, \techName{} achieves the highest F1 on all of them, outperforming multimodal graph baselines. 
This consistent advantage indicates that explicitly modeling transaction semantics and fund-flow propagation is effective across diverse dataset settings.
In contrast,~\ding{183}~\textbf{heuristic and traditional machine learning methods show unstable performance}, as fixed rules and shallow statistical features are insufficient to capture complex laundering behaviors such as cross-chain transfers, mixing, and multi-stage fund movements. ~\ding{184}~\textbf{Graph-based methods improve detection but still suffer from limited structural expressiveness.}~Graph-based methods improve detection by leveraging transaction graph structure, but their lower ranks indicate that structure alone cannot fully characterize diverse laundering strategies. Moreover, Table~\ref{tab:avg_f1_new} shows that~\ding{185}~\textbf{\techName{} maintains the most stable advantage in day-by-day detection}, achieving an average rank of 1.1 and ranking first on 12 out of 14 days.
In particular, during the late stage from Day~\circledc{12} to Day~\circledc{14}, many baselines degrade substantially. In contrast, by combining transaction semantics parsing with fund-flow modeling, \techName{} remains robust under sparse labels, class imbalance, and evolving laundering behaviors.

\begin{figure*}[t]
\centering
\subfloat[Time window~$\tau_t$]{
  \includegraphics[width=0.3\linewidth]{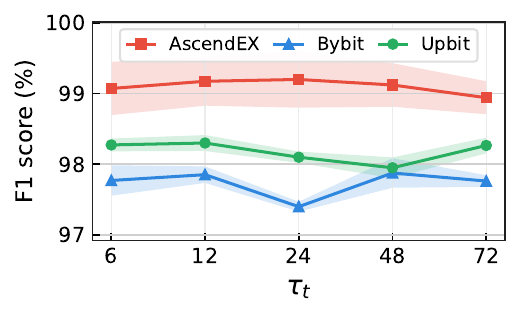}}
\hfill
\subfloat[Flow ratio~$\tau_f$]{
  \includegraphics[width=0.3\linewidth]{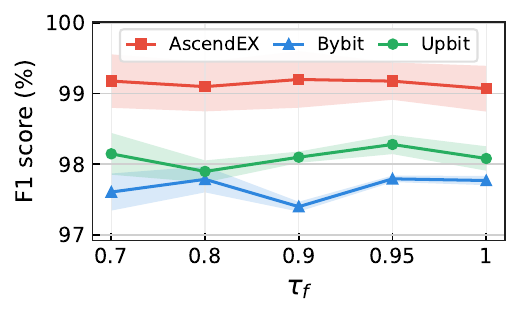}}
\hfill
\subfloat[Amount similarity~$\tau_a$]{
  \includegraphics[width=0.3\linewidth]{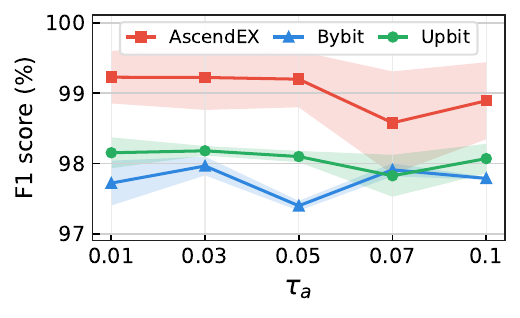}}
\caption{\label{fig:param_sensitivity}Parameter sensitivity of \techName{} under different fund-flow construction thresholds.}
\end{figure*}

\begin{figure*}[t]
    \centering
          \subfloat[Transaction distribution in \textit{BybitML}\label{fig:sum_daily}]{%
      \includegraphics[width=0.29\textwidth]{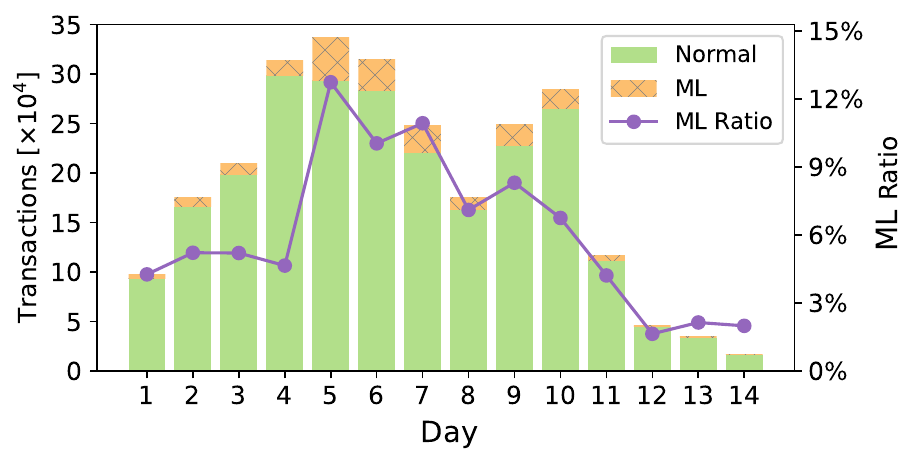}}%
    \hfill
      \subfloat[SAR quality\label{fig:sar_rate_sub}]{%
        \includegraphics[width=0.29\textwidth]{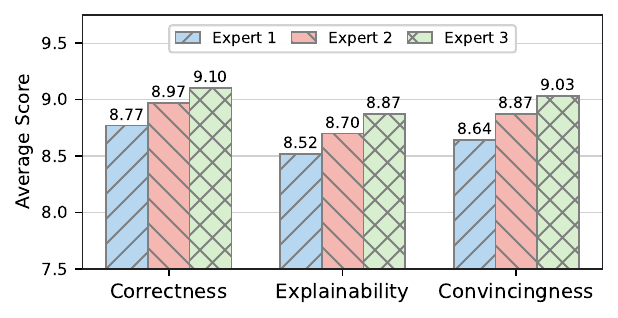}}%
    \hfill
          \subfloat[Detection time\label{fig:scalability_sub}]{%
      \includegraphics[width=0.29\textwidth]{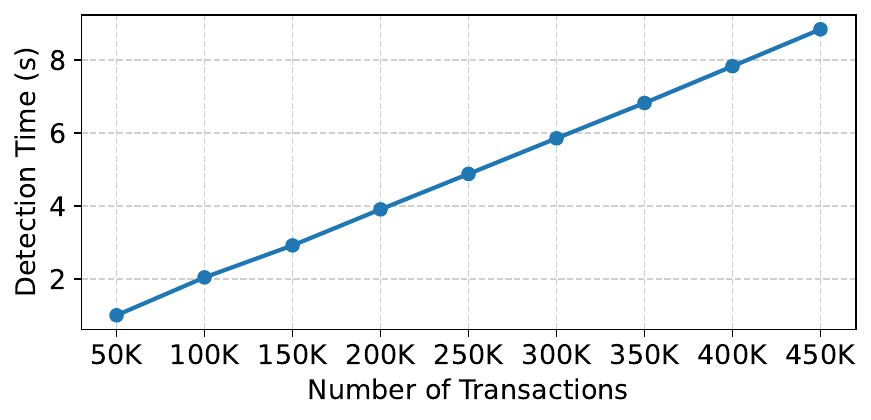}}%
    \hfill

    % \caption{\caption{Additional analyses for \textsc{FlowShield}: (a) daily transaction distribution in \textit{BybitML}, (b) expert-rated SAR quality, and (c) detection time under different transaction volumes.}zn \textit{BybitML}, (b) average SAR quality scores from expert ratings, and (c) detection time under increasing transaction volumes.}
    \caption{Additional analyses: (a) daily transaction distribution in \textit{BybitML}, (b) expert-rated SAR quality, and (c) detection scalability.}
    \label{fig:eval_three}
\end{figure*}

\begin{figure*}[t]
\centering
\subfloat[Semantic evolution.\label{fig:case_semantic}]{
  \includegraphics[width=0.26\linewidth]{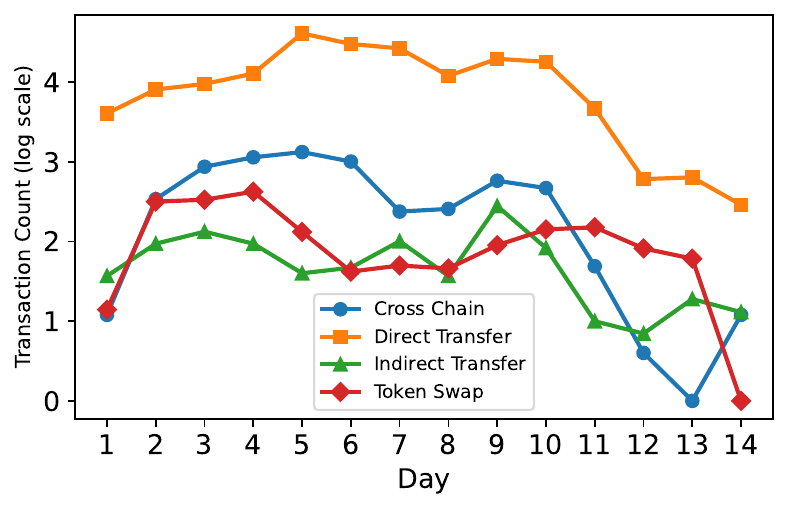}}
\hfill
\subfloat[Cumulative cross-chain volume.\label{fig:case_cross}]{
  \includegraphics[width=0.34\linewidth]{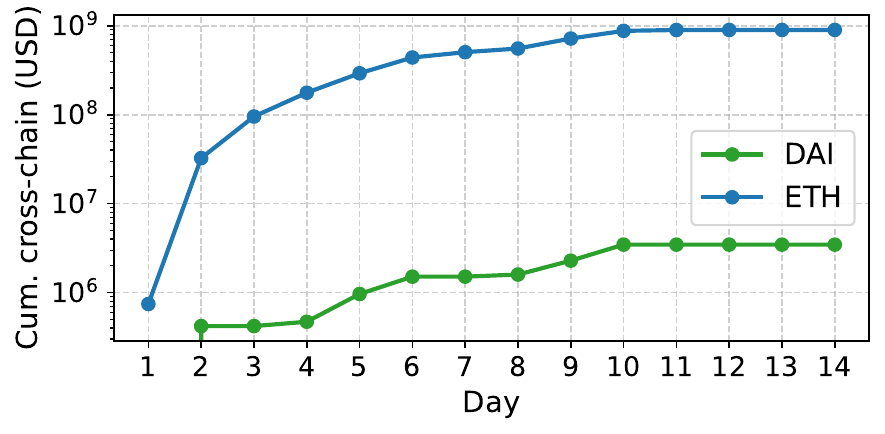}}
\hfill
\subfloat[Swap intensity.\label{fig:case_swap}]{
  \includegraphics[width=0.28\linewidth]{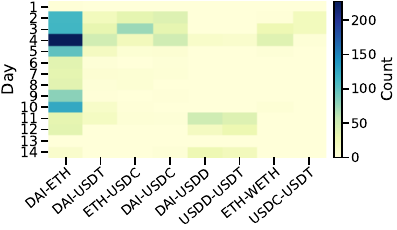}}
\caption{\label{fig:case_behavior} Behavioral insights from the Bybit hack incident.}
\end{figure*}

\subsection{Ablation Study~(RQ2)}

To validate the effectiveness of each component of \techName{}, we make corresponding modifications to \techName{} and design four ablated variants: 1) \textbf{w/o semantics}, which removes the semantics parsing module; 2) \textbf{w/o flow}, which removes the fund-flow construction, thus losing flow-aware structural context; 3) \textbf{w/o fusion}, which removes the bidirectional interaction layer and directly concatenates the projected text and structural representations; 4) \textbf{w/o Llama3}, which replaces Llama3 with BERT~\cite{devlin-etal-2019-bert} to encode transaction texts. The results are demonstrated in Table~\ref{tab:ablation_avg_new}. From the results, we can conclude that all the components significantly contribute to the performance. Among them, \ding{182}~\textbf{the semantics parsing module has a significant impact}, as w/o semantics leads to the largest F1 drop across all datasets. 
In addition, w/o flow also causes clear performance degradation, indicating that \ding{183}~\textbf{flow-aware structural context is important for capturing multi-hop laundering paths}. And w/o fusion and w/o Llama3 show the \ding{184}~\textbf{full text-structure fusion with Llama3 provides more stable and robust representations}. 
Notably, \textit{Bybit-BC} is the only exception where w/o fusion marginally higher than the full model. 
We attribute it to the UTXO-style Bitcoin data, where fund-flow relations are more explicit and contract-level semantic signals are less diverse than in account-based blockchains. Thus, simple concatenation can occasionally be sufficient on this subset, while the full fusion module provides more consistent gains on the other three datasets and the best average performance.

\input{Tabs/tab_ablation_avg_new}

% 为验证各模块的有效性，我们在四个数据集上进行消融实验，结果如 Table~\ref{tab:ablation_avg_new} 所示。可以观察到，完整的 \techName{} 在多数数据集上取得最优表现，说明语义解析、资金流建模、融合模块和 Llama3 表征共同提升了检测效果。

% ~\ding{182}~\textbf{语义解析是最关键的组成部分。}
% 移除 semantics 后，所有数据集上的 F1 均出现明显下降，尤其在 \textit{Bybit-EC}、\textit{UpbitHack} 和 \textit{AscendEX} 上分别降至 89.7\%、91.3\% 和 91.8\%。这说明仅依赖结构或统计特征难以区分语义相近但行为意图不同的交易。

% ~\ding{183}~\textbf{资金流建模与融合模块提供稳定增益。}
% 移除 flow 后，模型在多个数据集上仍有明显下降，表明资金流上下文对于捕获多阶段洗钱路径是必要的。相比之下，移除 fusion 或 Llama3 后性能下降较小，但整体仍低于完整模型，说明文本语义与图结构的联合建模能够进一步提升鲁棒性。

% ~\ding{184}~\textbf{完整模型整体最稳定。}
% \techName{} 在 \textit{Bybit-EC}、\textit{UpbitHack} 和 \textit{AscendEX} 上取得最高 F1，并在 \textit{Bybit-BC} 上保持与最优结果接近的性能。这表明各模块具有互补作用，使模型能够在不同链和不同交易场景下保持稳定表现。

% the complete \techName{} achieves the best performance, showing that all modules jointly contribute to effective detection. W/o semantics parsing significantly reduces both precision and recall, causing the largest F1 drop and highlighting the importance of semantics for characterizing transaction behaviors. W/o flow awareness, recall drops noticeably, indicating more missed laundering transactions when flow-based context is absent. W/o text-structure fusion, the model can still leverage semantic and structural features separately and maintain reasonable precision and recall, but remains inferior to the full system, confirming that text-structure interaction is vital.

% \subsection{Parameter Analysis}
\subsection{Sensitivity Analysis~(RQ3)}

\heading{Parameter Sensitivity.}
We analyze the sensitivity of \techName{} to three fund-flow construction parameters: the time window~$\tau_t$, flow ratio~$\tau_f$, and amount similarity~$\tau_a$.
As shown in Figure~\ref{fig:param_sensitivity}, \techName{} remains stable under different parameter settings. Changing $\tau_t$, $\tau_f$, or $\tau_a$ across the datasets only leads to small F1 fluctuations, and the performance consistently stays at a high level. This indicates that \techName{} does not rely on a fragile threshold choice. The default setting, $\tau_t=24$, $\tau_f=0.9$, and $\tau_a=0.05$, provides a balanced configuration: it preserves sufficient temporal and value-flow context while avoiding excessive irrelevant transactions in the constructed fund-flow subgraphs.

\heading{Sensitivity to Semantic Drift.}
We manually audited 400 parsed transaction groups, covering 100 cases per semantic type, and found all assignments consistent with the corresponding transfer-group relations, token-flow relations, and bridge logs. 
We then evaluate robustness to semantic omission, the main practical risk in our parser: true cross-chain transfers may be parsed as direct transfers when bridge lists or logs are incomplete. We simulate this drift on \textit{Bybit-EC} by randomly reverting 10\%--50\% of true cross-chain transfers to direct transfers. As shown in Table~\ref{tab:bybit_parser_robustness}, even at 50\% perturbation, overall F1 drops by only 0.03\%, while cross-chain F1 drops by at most 0.99\%. These results indicate that \techName{} is insensitive to moderate semantic drift, with fund-flow and structural signals compensating for semantic omissions.

\begin{table}[t]
\centering
% \caption{F1 drop under different perturbation ratios.}
\caption{F1 drop under different semantic-drift perturbation ratios.}
\label{tab:bybit_parser_robustness}
\begin{tabular}{lccccccc}
\toprule
$\Delta$ F1 (\%) & 0\% & 10\% & 20\% & 30\% & 40\% & 50\% & Avg. \\
\midrule
Overall     & 0.00 & 0.00 & 0.00 & 0.01 & 0.02 & 0.03 & 0.01 \\
Cross-chain & 0.00 & 0.19 & 0.23 & 0.46 & 0.72 & 0.99 & 0.52 \\
\bottomrule
\end{tabular}
\end{table}

\subsection{Detection Scalability~(RQ4)}

Figure~\ref{fig:scalability_sub} shows the scalability of the detection stage in \textsc{FlowShield}. As the number of transactions increases from 50K to 450K, the runtime time increases roughly linearly, and the average latency remains stable at ~0.02ms per transaction, far below on-chain confirmation time.
This demonstrates that the detection stage of \textsc{FlowShield} has low computational overhead for post-event investigation.

\section{Case Study}

\subsection{Behavior Analysis}

% \begin{figure*}[t]
% \centering
% \subfloat[Semantic evolution.\label{fig:case_semantic}]{
%   \includegraphics[width=0.26\linewidth]{Figs/semantic_tx_num.pdf}}
% \hfill
% \subfloat[Cross-chain amount.\label{fig:case_cross}]{
%   \includegraphics[width=0.34\linewidth]{Figs/cross_value.pdf}}
% \hfill
% \subfloat[Swap intensity.\label{fig:case_swap}]{
%   \includegraphics[width=0.28\linewidth]{Figs/swap_pair_activity_heatmap_compact.pdf}}
% \caption{\label{fig:case_behavior}Behavioral insights from the Bybit hack incident.}
% \end{figure*}

%我们对检测到的洗钱交易进行逐日行为分析，发现整体以直接转账为主，早期换币与跨链活动更活跃，体现出攻击者倾向于在案发初期通过资产转换与跨链迁移来提升追踪难度。换币行为以 DAI↔ETH 最为频繁，稳定币之间的多次互换进一步扰动资金路径；跨链转移主要经由 Thorchain 从以太坊迁移至比特币，接收端呈现“多数短命地址 + 少数聚合地址”的长尾结构。与此同时，规避策略上早期更偏向Automation（脚本化批量分散），跨链相关交易则更多呈现 **Transit（中转后全额转发）**模式，表明资金常通过中转层再进入桥以降低直接暴露风险；更完整的统计结果与案例细节见附录。
% \textsc{FlowShield} provides a case-level view of the laundering transactions detected in the \textit{Bybit hack}. As shown in Figure~\ref{fig:case_behavior}, it reveals three key behaviors in the multi-stage laundering process: ~\ding{182}~different transaction semantics are dynamically combined across days; ~\ding{183}~cross-chain transfers mainly move DAI and ETH from Ethereum to Bitcoin; and ~\ding{184}~high-liquidity swap pairs such as DAI--ETH are frequently used to obfuscate fund paths. These findings show that \textsc{FlowShield} not only detects suspicious transactions, but also provides behavioral insights for understanding complex laundering processes.

\textsc{FlowShield} provides a case-level view of the laundering transactions detected in the {Bybit hack}. 
As shown in Figure~\ref{fig:case_behavior}, it reveals three key behaviors in the multi-stage laundering process.~\ding{182}~Transaction semantics change dynamically across days, suggesting that launderers do not rely on a single operation but combine dispersal, swaps, and cross-chain movements as the laundering process evolves.~\ding{183}~Cross-chain transfers mainly move DAI and ETH from Ethereum to Bitcoin through Thorchain, showing that chain-hopping is used to break single-chain fund-flow continuity.~\ding{184}~High-liquidity swap pairs such as DAI--ETH are frequently used, indicating that launderers prefer liquid assets to change token forms while reducing execution friction. 
These findings show that \textsc{FlowShield} not only detects suspicious transactions, but also provides behavioral insights for understanding complex laundering processes.

\subsection{SARs Quality}

% \begin{figure}[t]
%   \centering
%   \includegraphics[width=0.4\textwidth]{Figs/eg_sar.pdf}
%   % \caption{\label{fig:sar_example}Example SAR generated by \techName{}, including a \textit{FlowDiagram}, \textit{Summary}, and \textit{RedFlags}.}
%   \caption{\label{fig:sar_example}Abridged SAR example generated by \techName{}, highlighting the \textit{FlowDiagram}, \textit{Summary}, and \textit{RedFlags}.}

% \end{figure}

% \textsc{FlowShield} further organizes suspicious fund flows into SARs with a flow diagram, a concise summary, and key red flags, converting behaviors such as automated dispersal, frequent swaps, and rapid forwarding into reviewable evidence. Figure~\ref{fig:sar_example} shows one generated SAR example. As shown in Figure~\ref{fig:sar_rate_sub}, ratings from blockchain security experts show that the generated SARs score above 8.5 in \textit{correctness}, \textit{explainability}, and \textit{convincingness}, indicating their practical utility for investigation.

\textsc{FlowShield} further organizes suspicious fund flows into SARs with a flow diagram, a concise summary, and key red flags, converting behaviors into reviewable investigation cues. Figure~\ref{fig:sar_example} shows one generated SAR example. 
To assess report quality, we randomly sample 50 SARs and invite three blockchain security experts to rate each report on a 1--10 scale. 
The evaluation covers \textit{correctness} (whether the report matches the underlying fund flow), \textit{explainability} (whether it helps understand the suspicious process), and \textit{convincingness} (whether the red flags support manual review). 
As shown in Figure~\ref{fig:sar_rate_sub}, the generated SARs score above 8.5 across all dimensions, indicating their practical utility for investigation.

\section{Conclusion}
We presented \textsc{FlowShield}, a Crypto AML framework for detecting complex laundering transactions and generating investigator-facing suspicious activity reports. 
\textsc{FlowShield} jointly models transaction semantics, localized fund-flow context, and graph-structural information, enabling it to characterize laundering behaviors beyond isolated direct transfers. 
To support multi-chain Crypto AML research, we constructed \textit{BybitML}, the first public multi-chain laundering dataset derived from a real-world security incident. 
Comprehensive experiments on \textit{BybitML} and two public datasets show that \textsc{FlowShield} consistently outperforms representative baselines and maintains stable day-by-day performance. 
Further behavior and SAR analyses demonstrate that \textsc{FlowShield} can reveal diverse laundering strategies and produce readable reports for investigating complex multi-hop fund flows.

\bibliographystyle{IEEEtran}
\bibliography{my_ref}

\end{document}

%% file: chapter/Alg-flow.tex
\begin{algorithm}[t]
\caption{Fund-flow Construction}
% \scriptsize
\small
\DontPrintSemicolon
\label{alg:flow_awareness}

\KwIn{Transaction MultiDiGraph $G$; target transaction $t_{tg}$; flow template $\mathsf{Tmp}_{f}$; thresholds $\tau_t$, $\tau_f$, $\tau_a$}
\KwOut{Fund-flow subgraphs $SubG_{up}$, $SubG_{down}$, $SubG_{para}$; flow pattern $\mathsf{T}_f$; flow text $\mathsf{Txt}_{flow}$}

$(SubG_{up}, \text{Tx}_{up}) \leftarrow \mathrm{ExtractUpstream}(G,t_{tg},\tau_t,\tau_f)$\;
$(SubG_{down}, \text{Tx}_{down}) \leftarrow \mathrm{ExtractDown}(G,t_{tg},\tau_t,\tau_f)$\;
$(SubG_{para}, \text{Tx}_{para}) \leftarrow \mathrm{ExtractParallel}(G,t_{tg},\tau_t,\tau_a)$\;

$A_{down} \leftarrow \sum_{e \in \text{Tx}_{down}} e(\text{amount})$\;

\uIf{$|\text{Tx}_{down}| = 0$}{
    $\mathsf{T}_f \leftarrow$ \textit{No forward}\;
}
\uElseIf{$|\text{Tx}_{para}| > 0$ \textbf{and} $A_{down} \ge t_{tg}(\text{amount})$}{
    $\mathsf{T}_f \leftarrow$ \textit{Mix}\;
}
\uElseIf{$A_{down} \ge \tau_f \cdot t_{tg}(\text{amount})$}{
    \uIf{$|\text{Tx}_{down}| = 1$}{
        $\mathsf{T}_f \leftarrow$ \textit{Single outflow}\;
    }
    \Else{
        $\mathsf{T}_f \leftarrow$ \textit{Entire disperse}\;
    }
}
\ElseIf{$|\text{Tx}_{down}| > 1$}{
    $\mathsf{T}_f \leftarrow$ \textit{Partial disperse}\;
}

$\mathsf{Txt}_{flow} \leftarrow \mathrm{Verbalize}(\mathsf{Tmp}_{f}, SubG_{up}, SubG_{down}, SubG_{para}, \mathsf{T}_f)$\;

\end{algorithm}

%% file: Tabs/tab_effectiveness_avg_new.tex
\begin{table}[t]
\small
\centering
\caption{F1 scores (\%) of different methods on four datasets. Highlighted are the results ranked \textcolor{red}{\underline{first}}, \textcolor{green!50!black}{\underline{second}}, and \textcolor{blue}{\underline{third}}. Rank denotes the average ranking over 4 datasets.}
\label{tab:avg_metrics_new}
\resizebox{\columnwidth}{!}{
\begin{tabular}{l c c c c c}
\toprule
\textbf{Method} & \textbf{\textit{Bybit-EC}} & \textbf{\textit{Bybit-BC}} & \textbf{\textit{Upbit}} & \textbf{\textit{AscendEX}} & Rank \\
\midrule
\rowcolor{gray!20}
\multicolumn{6}{c}{Heuristic Methods} \\
Cyclic~\cite{lv2023detection} & 6.9{\scriptsize$\pm$0.0} & 0.7{\scriptsize$\pm$0.0} & 4.4{\scriptsize$\pm$0.0} & 2.1{\scriptsize$\pm$0.0} & 12.8 \\
XBlock~\cite{wu2023towards} & 16.7{\scriptsize$\pm$0.0} & -- & -- & -- & 11.0 \\
\midrule
\rowcolor{gray!20}
\multicolumn{6}{c}{Machine learning Methods} \\
MLP~\cite{rosenblatt1958perceptron} & 9.2{\scriptsize$\pm$10.4} & 34.4{\scriptsize$\pm$18.3} & 30.6{\scriptsize$\pm$15.0} & 15.9{\scriptsize$\pm$7.0} & 11.0 \\
LR~\cite{cox1958regression} & 61.3{\scriptsize$\pm$0.6} & 68.4{\scriptsize$\pm$0.5} & 34.4{\scriptsize$\pm$0.9} & 10.4{\scriptsize$\pm$0.2} & 9.3 \\
SVM~\cite{cortes1995support} & 61.2{\scriptsize$\pm$0.66} & 69.6{\scriptsize$\pm$0.5} & 32.8{\scriptsize$\pm$1.1} & 9.5{\scriptsize$\pm$0.4} & 9.8 \\
\midrule
\rowcolor{gray!20}
\multicolumn{6}{c}{Graph Neural Network Methods} \\
GCN~\cite{kipf2016semi} & 32.3{\scriptsize$\pm$16.2} & 82.5{\scriptsize$\pm$0.8} & 88.2{\scriptsize$\pm$1.2} & 78.0{\scriptsize$\pm$3.1} & 7.0 \\
SAGE~\cite{hamilton2017inductive} & 72.1{\scriptsize$\pm$16.1} & 89.2{\scriptsize$\pm$1.9} & \textcolor{green!50!black}{\underline{97.8{\scriptsize$\pm$0.2}}} & 95.3{\scriptsize$\pm$0.4} & 3.8 \\
MRGNN~\cite{hyun2023anti} & 81.3{\scriptsize$\pm$4.0} & 89.0{\scriptsize$\pm$1.3} & 88.7{\scriptsize$\pm$0.8} & 92.9{\scriptsize$\pm$1.5} & 4.8 \\
\midrule
\rowcolor{gray!20}
\multicolumn{6}{c}{Graph Pattern Mining Methods} \\
Tracer~\cite{wu2023tracer} & 21.9{\scriptsize$\pm$0.0} & -- & 62.5{\scriptsize$\pm$0.0} & 43.7{\scriptsize$\pm$0.0} & 8.7 \\
DenseFlow~\cite{lin2024denseflow} & 9.5{\scriptsize$\pm$0.0} & 0.7{\scriptsize$\pm$0.0} & 59.5{\scriptsize$\pm$0.0} & 12.5{\scriptsize$\pm$0.0} & 10.5 \\
MPOCryptoML~\cite{samadi2025mpocryptoml} & 71.3{\scriptsize$\pm$3.2} & 77.9{\scriptsize$\pm$1.6} & 84.2{\scriptsize$\pm$2.0} & 79.4{\scriptsize$\pm$14.0} & 6.5 \\
\midrule
\rowcolor{gray!20}
\multicolumn{6}{c}{Multimodal Graph Methods} \\
GraphGPS~\cite{rampavsek2022recipe} & \textcolor{green!50!black}{\underline{96.0{\scriptsize$\pm$4.0}}} & \textcolor{green!50!black}{\underline{94.7{\scriptsize$\pm$0.2}}} & \textcolor{blue}{\underline{96.5{\scriptsize$\pm$0.3}}} & \textcolor{green!50!black}{\underline{97.6{\scriptsize$\pm$0.4}}} & \textcolor{green!50!black}{\underline{2.3}} \\
GraphFormers~\cite{yang2021graphformers} & \textcolor{blue}{\underline{95.7{\scriptsize$\pm$5.3}}} & \textcolor{blue}{\underline{94.4{\scriptsize$\pm$0.5}}} & 96.4{\scriptsize$\pm$0.3} & \textcolor{blue}{\underline{97.5{\scriptsize$\pm$0.5}}} & \textcolor{blue}{\underline{3.3}} \\
\midrule
\rowcolor{gray!20}
\multicolumn{6}{c}{Ours} \\
\textbf{FlowShield} & \textcolor{red}{\underline{\textbf{97.2{\scriptsize$\pm$3.0}}}} & \textcolor{red}{\underline{\textbf{97.4{\scriptsize$\pm$0.07}}}} & \textcolor{red}{\underline{\textbf{98.1{\scriptsize$\pm$0.08}}}} & \textcolor{red}{\underline{\textbf{99.2{\scriptsize$\pm$0.4}}}} & \textcolor{red}{\underline{\textbf{1.0}}} \\
\bottomrule
\end{tabular}
}
\par\vspace{1mm}
\noindent
\begin{minipage}{\columnwidth}
\footnotesize\raggedright
XBlock and Tracer are not designed for the Bitcoin blockchain, and therefore cannot detect \textit{Bybit-BC}.
Since the labels of \textit{Upbit} and \textit{AscendEX} are collected using XBlock, XBlock is excluded from the evaluation on these two datasets.
\end{minipage}
\end{table}

%% file: Tabs/tab_comparative_exp_new.tex
\begin{table*}[t]
\centering
%Effectiveness comparison with different AML methods on Bybit Hack datasets.
\caption{F1 scores (\%) of different methods on \textit{Bybit-EC}.   \circledc{1} to \circledc{14} denote the data of \textit{Bybit-EC} from the first day to the fourteenth day. Highlighted are the results ranked \textcolor{red}{\underline{first}}, \textcolor{green!50!black}{\underline{second}}, and \textcolor{blue}{\underline{third}}. Rank denotes the average ranking over 14 days.}
\label{tab:avg_f1_new}
\resizebox{\textwidth}{!}{
\begin{tabular}{l c c c c c c c c c c c c c c c}
\toprule
\textbf{Methods} & \circled{1} & \circled{2} & \circled{3} & \circled{4} & \circled{5} & \circled{6} & \circled{7} & \circled{8} & \circled{9} & \circled{10} & \circled{11} & \circled{12} & \circled{13} & \circled{14} & Rank \\
\midrule
\rowcolor{gray!20}
\multicolumn{16}{c}{Heuristic Methods} \\
Cyclic~\cite{lv2023detection}     & 4.7 & 6.7 & 6.9 & 6.6 & 11.4 & 9.9 & 9.9 & 8.3 & 9.3 & 8.1 & 5.6 & 2.7 & 3.4 & 2.7 & 12.6 \\
XBlock~\cite{wu2023towards}     & 35.2 & 31.3 & 38.4 & 35.8 & 16.9 & 18.8 & 10.4 & 17.7 & 24.5 & 0.9 & 2.1 & 0.5 & 0.3 & 1.2 & 10.9 \\
\midrule
\rowcolor{gray!20}
\multicolumn{16}{c}{Machine learning Methods} \\
MLP~\cite{rosenblatt1958perceptron}        & 7.7 & 15.4 & 7.5 & 7.7 & 8.2 & 13.3 & 8.2 & 9.4 & 7.1 & 3.8 & 13.1 & 11.1 & 13.1 & 3.3 & 11.9 \\
LR~\cite{cox1958regression}         & 71.6 & 77.1 & 67.0 & 70.3 & 87.7 & 87.1 & 88.1 & 66.9 & 80.2 & 82.9 & 79.2 & 0.0 & 0.0 & 0.0 & 7.9 \\
SVM~\cite{cortes1995support}        & 71.6 & 77.6 & 69.8 & 69.9 & 86.6 & 86.8 & 88.0 & 67.0 & 78.4 & 83.2 & 77.4 & 0.0 & 0.0 & 0.0 & 8.1 \\
\midrule
\rowcolor{gray!20}
\multicolumn{16}{c}{Graph Neural Network Methods} \\
GCN~\cite{kipf2016semi}        & 15.0 & 9.8 & 0.9 & 8.1 & 14.6 & 22.9 & 90.8 & 44.8 & 60.8 & 39.1 & 60.5 & 58.6 & 17.8 & 7.8 & 9.2 \\
SAGE~\cite{hamilton2017inductive}  & 48.4 & 67.5 & 64.2 & 76.9 & 93.7 & 88.6 & 92.5 & 87.7 & 83.0 & 82.9 & 80.2 & 50.9 & 40.4 & 52.3 & 5.6 \\
MRGNN~\cite{hyun2023anti} & 91.2 & 85.5 & 85.1 & 87.9 & 97.2 & 96.0 & 97.2 & 93.4 & 96.5 & 89.9 & 82.7 & 42.0 & 49.2 & 44.6 & 4.4 \\
\midrule
\rowcolor{gray!20}
\multicolumn{16}{c}{Graph Pattern Mining Methods} \\
Tracer~\cite{wu2023tracer}     & 19.1 & 31.6 & 36.8 & 27.7 & 14.0 & 16.4 & 10.0 & 18.5 & 32.3 & 23.9 & 20.2 & 21.3 & 25.9 & 8.8 & 9.6 \\
DenseFlow~\cite{lin2024denseflow} & 5.0 & 7.3 & 9.8 & 5.5 & 8.4 & 9.9 & 19.5 & 13.0 & 15.0 & 9.4 & 7.9 & 3.2 & 3.1 & 3.9 & 11.7 \\
MPOCryptoML~\cite{samadi2025mpocryptoml} & 60.0 & 66.3 & 68.4 & 75.5 & 88.9 & 87.1 & 90.3 & 80.3 & 80.0 & 82.5 & 72.0 & 48.0 & 52.9 & 44.5 & 6.5 \\
\midrule
\rowcolor{gray!20}
\multicolumn{16}{c}{Multimodal Graph Methods} \\
GraphGPS~\cite{rampavsek2022recipe} & \textcolor{green!50!black}{\underline{97.2}} & \textcolor{blue}{\underline{94.9}} & \textcolor{green!50!black}{\underline{95.9}} & \textcolor{blue}{\underline{97.4}} & \textcolor{red}{\underline{\textbf{99.6}}} & \textcolor{green!50!black}{\underline{99.3}} & \textcolor{blue}{\underline{99.4}} & \textcolor{blue}{\underline{98.6}} & \textcolor{blue}{\underline{99.0}} & \textcolor{green!50!black}{\underline{98.6}} & \textcolor{blue}{\underline{97.2}} & \textcolor{blue}{\underline{88.4}} & \textcolor{green!50!black}{\underline{89.8}} & \textcolor{blue}{\underline{88.9}} & \textcolor{blue}{\underline{2.5}} \\
GraphFormers~\cite{yang2021graphformers} & \textcolor{blue}{\underline{96.9}} & \textcolor{green!50!black}{\underline{96.8}} & \textcolor{blue}{\underline{94.5}} & \textcolor{green!50!black}{\underline{97.9}} & \textcolor{blue}{\underline{99.2}} & \textcolor{green!50!black}{\underline{99.3}} & \textcolor{red}{\underline{\textbf{99.6}}} & \textcolor{green!50!black}{\underline{99.2}} & \textcolor{red}{\underline{\textbf{99.4}}} & \textcolor{blue}{\underline{98.1}} & \textcolor{red}{\underline{\textbf{98.1}}} & \textcolor{green!50!black}{\underline{90.8}} & \textcolor{blue}{\underline{80.7}} & \textcolor{green!50!black}{\underline{90.1}} & \textcolor{green!50!black}{\underline{2.1}} \\
\midrule
\rowcolor{gray!20}
\multicolumn{16}{c}{Ours} \\
\textbf{FlowShield} & \textcolor{red}{\underline{\textbf{98.3}}} & \textcolor{red}{\underline{\textbf{97.9}}} & \textcolor{red}{\underline{\textbf{97.8}}} & \textcolor{red}{\underline{\textbf{98.7}}} & \textcolor{green!50!black}{\underline{99.4}} & \textcolor{red}{\underline{\textbf{99.5}}} & \textcolor{red}{\underline{\textbf{99.6}}} & \textcolor{red}{\underline{\textbf{99.3}}} & \textcolor{red}{\underline{\textbf{99.4}}} & \textcolor{red}{\underline{\textbf{99.0}}} & \textcolor{green!50!black}{\underline{97.6}} & \textcolor{red}{\underline{\textbf{93.1}}} & \textcolor{red}{\underline{\textbf{90.2}}} & \textcolor{red}{\underline{\textbf{93.1}}} & \textcolor{red}{\underline{\textbf{1.1}}} \\
\bottomrule
\end{tabular}
}
\end{table*}

%% file: Tabs/tab_dataset_stats.tex
\begin{table}[t]
\caption{Statistics of the datasets.}
\small
\centering
\label{tab:dataset_stats}
\resizebox{\columnwidth}{!}{
\begin{tabular}{l c c c c}
\toprule
\textbf{Metric} & \textbf{\textit{Bybit-EC}} & \textbf{\textit{Bybit-BC}} & \textbf{\textit{Upbit}} & \textbf{\textit{AscendEX}} \\
\midrule
\#Node~(address) & 58,790 & 62,409 & 16,397 & 6,713 \\
\#Edge~(transaction) & 2,616,736 & 134,280 & 40,287 & 29,074 \\
\#ML~(laundering transaction) & 199,518 & 31,176 & 17,488 & 1,940 \\
\%ML~(laundering ratio)  & 7.6 & 23.2 & 43.4 & 6.7 \\
\bottomrule
\end{tabular}
}
\end{table}

%% file: Tabs/tab_ablation_avg_new.tex
\begin{table}[t]
\small
\centering
\caption{Ablation study results in F1 score (\%) on four datasets. Highlighted are the results ranked \textcolor{red}{first}, \textcolor{green!50!black}{second}, and \textcolor{blue}{third}.}
\label{tab:ablation_avg_new}
\resizebox{\columnwidth}{!}{
\begin{tabular}{l c c c c}
\toprule
\textbf{Method} & \textbf{\textit{Bybit-EC}} & \textbf{\textit{Bybit-BC}} & \textbf{\textit{Upbit}} & \textbf{\textit{AscendEX}} \\
\midrule
w/o semantics & 89.7{\scriptsize$\pm$2.4} & 91.9{\scriptsize$\pm$0.6} & 91.3{\scriptsize$\pm$1.0} & 91.8{\scriptsize$\pm$1.8} \\
w/o flow & 94.4{\scriptsize$\pm$2.3} & 92.3{\scriptsize$\pm$0.3} & 94.0{\scriptsize$\pm$0.3} & 98.3{\scriptsize$\pm$0.4} \\
w/o fusion & \textcolor{blue}{95.9{\scriptsize$\pm$1.5}} & \textcolor{red}{\textbf{97.6{\scriptsize$\pm$0.2}}} & \textcolor{green!50!black}{97.4{\scriptsize$\pm$0.4}} & \textcolor{blue}{98.7{\scriptsize$\pm$0.3}} \\
w/o Llama3 & \textcolor{green!50!black}{96.5{\scriptsize$\pm$3.8}} & \textcolor{green!50!black}{97.4{\scriptsize$\pm$0.2}} & \textcolor{blue}{97.3{\scriptsize$\pm$0.1}} & \textcolor{green!50!black}{98.8{\scriptsize$\pm$0.3}} \\
\textbf{FlowShield} & \textcolor{red}{\textbf{97.2{\scriptsize$\pm$3.0}}} & \textcolor{green!50!black}{{97.4{\scriptsize$\pm$0.07}}} & \textcolor{red}{\textbf{98.1{\scriptsize$\pm$0.08}}} & \textcolor{red}{\textbf{99.2{\scriptsize$\pm$0.4}}} \\
\bottomrule
\end{tabular}
}
\end{table}

%% file: main.bbl
\begin{thebibliography}{10}
\providecommand{\url}[1]{#1}
\csname url@samestyle\endcsname
\providecommand{\newblock}{\relax}
\providecommand{\bibinfo}[2]{#2}
\providecommand{\BIBentrySTDinterwordspacing}{\spaceskip=0pt\relax}
\providecommand{\BIBentryALTinterwordstretchfactor}{4}
\providecommand{\BIBentryALTinterwordspacing}{\spaceskip=\fontdimen2\font plus
\BIBentryALTinterwordstretchfactor\fontdimen3\font minus
  \fontdimen4\font\relax}
\providecommand{\BIBforeignlanguage}[2]{{%
\expandafter\ifx\csname l@#1\endcsname\relax
\typeout{** WARNING: IEEEtran.bst: No hyphenation pattern has been}%
\typeout{** loaded for the language `#1'. Using the pattern for}%
\typeout{** the default language instead.}%
\else
\language=\csname l@#1\endcsname
\fi
#2}}
\providecommand{\BIBdecl}{\relax}
\BIBdecl

\bibitem{campbell2018bitcoin}
M.~Campbell-Verduyn, ``Bitcoin, crypto-coins, and global anti-money laundering
  governance,'' \emph{Crime, Law and Social Change}, vol.~69, no.~2, pp.
  283--305, 2018.

\bibitem{26Crime}
TRM, ``2026 crypto crime report,'' 2026,
  https://www.trmlabs.com/reports-and-whitepapers/2026-crypto-crime-report.

\bibitem{report}
SlowMist, ``Blockchain security and {AML} annual report,'' 2024,
  https://www.slowmist.com/report/2024-Blockchain-Security-and-AML-Annual-Report(EN).pdf.

\bibitem{CASTILLOLEON2026102916}
J.~{Castillo León} and A.~Lehar, ``What data have told us about decentralized
  finance,'' \emph{Journal of Corporate Finance}, vol.~96, p. 102916, 2026.

\bibitem{ustreasury2023_defi_risk}
\BIBentryALTinterwordspacing
{U.S. Department of the Treasury}, ``Illicit finance risk assessment of
  decentralized finance,'' U.S. Department of the Treasury, Tech. Rep., 2023.
  [Online]. Available:
  \url{https://home.treasury.gov/system/files/136/DeFi-Risk-Full-Review.pdf}
\BIBentrySTDinterwordspacing

\bibitem{sok25fu}
Q.~Fu, J.~K. Liu, S.~Pan, and T.~H. Yuen, ``Sok: A deep dive into anti-money
  laundering techniques for blockchain cryptocurrencies,'' in
  \emph{Australasian Conference on Information Security and Privacy
  (ACISP)}.\hskip 1em plus 0.5em minus 0.4em\relax Springer, 2025, pp.
  310--330.

\bibitem{lv2023detection}
W.~Lv, J.~Liu, and L.~Zhou, ``Detection of money laundering address over the
  {E}thereum blockchain,'' in \emph{International Conference on Frontiers
  Technology of Information and Computer}.\hskip 1em plus 0.5em minus
  0.4em\relax Tsingtao, China: IEEE, 2023, pp. 866--869.

\bibitem{wu2023towards}
J.~Wu, D.~Lin, Q.~Fu, S.~Yang, T.~Chen, Z.~Zheng, and B.~Song, ``Towards
  understanding asset flows in crypto money laundering through the lenses of
  {E}thereum heists,'' \emph{IEEE Transactions on Information Forensics and
  Security (TIFS)}, vol.~19, pp. 1994--2009, 2023.

\bibitem{elmougy2023demystifying}
Y.~Elmougy and L.~Liu, ``Demystifying fraudulent transactions and illicit nodes
  in the {B}itcoin network for financial forensics,'' in \emph{ACM SIGKDD
  Conference on Knowledge Discovery and Data Mining (KDD)}.\hskip 1em plus
  0.5em minus 0.4em\relax Long Beach, CA, USA: ACM, 2023, pp. 3979--3990.

\bibitem{wang2024graphalm}
Q.~Wang, W.-T. Tsai, and T.~Shi, ``{GraphALM}: Active learning for detecting
  money laundering transactions on blockchain networks,'' \emph{IEEE Network},
  vol.~39, no.~2, pp. 294--303, 2025.

\bibitem{zhou2023visual}
F.~Zhou, Y.~Chen, C.~Zhu, L.~Jiang, X.~Liao, Z.~Zhong, X.~Chen, Y.~Chen, and
  Y.~Zhao, ``Visual analysis of money laundering in cryptocurrency exchange,''
  \emph{IEEE Transactions on Computational Social Systems (TCSS)}, vol.~11,
  no.~1, pp. 731--745, 2023.

\bibitem{alarab2020competence}
I.~Alarab, S.~Prakoonwit, and M.~I. Nacer, ``Competence of graph convolutional
  networks for anti-money laundering in bitcoin blockchain,'' in
  \emph{International Conference on Machine Learning Technologies
  (ICML)}.\hskip 1em plus 0.5em minus 0.4em\relax New York, NY, USA: ACM, 2020,
  pp. 23--27.

\bibitem{lin2024denseflow}
D.~Lin, J.~Wu, Y.~Yu, Q.~Fu, Z.~Zheng, and C.~Yang, ``Denseflow: Spotting
  cryptocurrency money laundering in {E}thereum transaction graphs,'' in
  \emph{Proceedings of the ACM on Web Conference (WWW)}.\hskip 1em plus 0.5em
  minus 0.4em\relax Singapore, Singapore: ACM, 2024, pp. 4429--4438.

\bibitem{nicholls2023fraudlens}
J.~Nicholls, A.~Kuppa, and N.-A. Le-Khac, ``Fraudlens: Graph structural
  learning for {B}itcoin illicit activity identification,'' in \emph{Annual
  Computer Security Applications Conference (ACSAC)}.\hskip 1em plus 0.5em
  minus 0.4em\relax Austin, TX, USA: ACM, 2023, pp. 324--336.

\bibitem{weber2019anti}
M.~Weber, G.~Domeniconi, J.~Chen, D.~K.~I. Weidele, C.~Bellei, T.~Robinson, and
  C.~Leiserson, ``Anti-money laundering in {B}itcoin: Experimenting with graph
  convolutional networks for financial forensics,'' in \emph{ACM SIGKDD
  International Conference on Knowledge Discovery and Data Mining
  (SIGKDD)}.\hskip 1em plus 0.5em minus 0.4em\relax Anchorage, AK, USA: ACM,
  2019.

\bibitem{wu2023tracer}
Z.~Wu, J.~Liu, J.~Wu, Z.~Zheng, and T.~Chen, ``Tracer: Scalable graph-based
  transaction tracing for account-based blockchain trading systems,''
  \emph{IEEE Transactions on Information Forensics and Security (TIFS)},
  vol.~18, pp. 2609--2621, 2023.

\bibitem{samadi2025mpocryptoml}
Y.~Samadi, H.~Dong, and X.~Xia, ``Mpocryptoml: Multi-pattern based off-chain
  crypto money laundering detection,'' \emph{arXiv preprint arXiv:2508.12641},
  2025.

\bibitem{hyun2023anti}
W.~Hyun, J.~Lee, and B.~Suh, ``Anti-money laundering in cryptocurrency via
  multi-relational graph neural network,'' in \emph{Pacific-Asia Conference on
  Knowledge Discovery and Data Mining (PAKDD)}, 2023, pp. 118--130.

\bibitem{kuccuk2023investigation}
D.~K{\"u}{\c{c}}{\"u}k, E.~{\c{C}}akar, {\"O}.~F. Yakut, and F.~Ertam,
  ``Investigation of cryptocurrency-centered money laundering scenarios in
  terms of digital forensics,'' \emph{International Journal of Advances in
  Engineering and Pure Sciences}, vol.~35, no.~3, pp. 285--296, 2023.

\bibitem{kipf2016semi}
T.~N. Kipf and M.~Welling, ``Semi-supervised classification with graph
  convolutional networks,'' \emph{arXiv preprint arXiv:1609.02907}, 2016.

\bibitem{hamilton2017inductive}
W.~Hamilton, Z.~Ying, and J.~Leskovec, ``Inductive representation learning on
  large graphs,'' \emph{Advances in Neural Information Processing Systems},
  vol.~30, p. 1025–1035, 2017.

\bibitem{grattafiori2024llama}
A.~Grattafiori, A.~Dubey, A.~Jauhri, A.~Pandey, A.~Kadian, A.~Al-Dahle,
  A.~Letman, A.~Mathur, A.~Schelten, A.~Vaughan \emph{et~al.}, ``The {Llama} 3
  herd of models,'' \emph{arXiv preprint arXiv:2407.21783}, 2024.

\bibitem{wei2021finetuned}
J.~Wei, M.~Bosma, V.~Y. Zhao, K.~Guu, A.~W. Yu, B.~Lester, N.~Du, A.~M. Dai,
  and Q.~V. Le, ``Finetuned language models are zero-shot learners,''
  \emph{arXiv preprint arXiv:2109.01652}, 2021.

\bibitem{hu2022lora}
E.~J. Hu, Y.~Shen, P.~Wallis, Z.~Allen-Zhu, Y.~Li, S.~Wang, L.~Wang, W.~Chen
  \emph{et~al.}, ``Lora: Low-rank adaptation of large language models,''
  \emph{arXiv preprint arXiv:2106.09685}, 2021.

\bibitem{tolstikhin2021mlp}
I.~O. Tolstikhin, N.~Houlsby, A.~Kolesnikov, L.~Beyer, X.~Zhai, T.~Unterthiner,
  J.~Yung, A.~Steiner, D.~Keysers, J.~Uszkoreit \emph{et~al.}, ``{MLP-mixer}:
  An all-mlp architecture for vision,'' \emph{Advances in Neural Information
  Processing Systems}, vol.~34, pp. 24\,261--24\,272, 2021.

\bibitem{shahriar2024putting}
S.~Shahriar, B.~D. Lund, N.~R. Mannuru, M.~A. Arshad, K.~Hayawi, R.~V.~K.
  Bevara, A.~Mannuru, and L.~Batool, ``Putting {GPT-4o} to the sword: {A}
  comprehensive evaluation of language, vision, speech, and multimodal
  proficiency,'' \emph{Applied Sciences}, vol.~14, no.~17, p. 7782, 2024.

\bibitem{rosenblatt1958perceptron}
F.~Rosenblatt, ``The {P}erceptron: {A} probabilistic model for information
  storage and organization in the brain.'' \emph{Psychological review},
  vol.~65, no.~6, p. 386, 1958.

\bibitem{cox1958regression}
D.~R. Cox, ``The regression analysis of binary sequences,'' \emph{Journal of
  the Royal Statistical Society Series B: Statistical Methodology}, vol.~20,
  no.~2, pp. 215--232, 1958.

\bibitem{cortes1995support}
C.~Cortes and V.~Vapnik, ``Support-vector networks,'' \emph{Machine learning},
  vol.~20, no.~3, pp. 273--297, 1995.

\bibitem{rampavsek2022recipe}
L.~Ramp{\'a}{\v{s}}ek, M.~Galkin, V.~P. Dwivedi, A.~T. Luu, G.~Wolf, and
  D.~Beaini, ``Recipe for a general, powerful, scalable graph transformer,''
  \emph{Advances in Neural Information Processing Systems}, vol.~35, pp.
  14\,501--14\,515, 2022.

\bibitem{yang2021graphformers}
J.~Yang, Z.~Liu, S.~Xiao, C.~Li, D.~Lian, S.~Agrawal, A.~Singh, G.~Sun, and
  X.~Xie, ``Graphformers: Gnn-nested transformers for representation learning
  on textual graph,'' \emph{Advances in Neural Information Processing Systems},
  vol.~34, pp. 28\,798--28\,810, 2021.

\bibitem{bybit25}
T.~Rajic and J.~Brock, ``The {B}ybit heist and the future of u.s. crypto
  regulation,'' 2025,
  https://www.csis.org/analysis/bybit-heist-and-future-us-crypto-regulation.

\bibitem{tokenview_blockchain_api}
{Tokenview}, ``Blockchain apis and data service platform,'' 2026,
  https://services.tokenview.io/en/product/api/op.

\bibitem{blockstream_explorer_api}
{Blockstream}, ``Blockstream explorer api,'' 2026,
  https://blockstream.info/explorer-api.

\bibitem{elliptic_bybit_six_months}
{Elliptic}, ``Bybit exploit: Six months on,'' 2025,
  https://www.elliptic.co/blog/bybit-exploit-six-months-on.

\bibitem{elliptic_data_fabric}
------, ``Elliptic data fabric,'' 2025,
  https://www.elliptic.co/platform/data-fabric.

\bibitem{arkham_intel_platform}
{Arkham Intelligence}, ``Arkham intel: Intel platform,'' 2025,
  https://intel.arkm.com/.

\bibitem{devlin-etal-2019-bert}
J.~Devlin, M.-W. Chang, K.~Lee, and K.~Toutanova, ``{BERT}: Pre-training of
  deep bidirectional transformers for language understanding,'' in
  \emph{Proceedings of the 2019 Conference of the North {A}merican Chapter of
  the Association for Computational Linguistics}, Minneapolis, Minnesota, Jun.
  2019, pp. 4171--4186.

\end{thebibliography}
